\documentclass[prd,nofootinbib,preprint,superscriptaddress]{revtex4}

\usepackage{graphicx}
\usepackage{amsmath}
\usepackage{amssymb}

\usepackage[hypertex]{hyperref}

\newcommand{\be}{\begin{equation}}
\newcommand{\ee}{\end{equation}}
\newcommand{\unit}[1]{\,\mathrm{#1}}
\newcommand{\vect}[1]{\boldsymbol{\rm #1}}
\newcommand{\bra}[1]{\ensuremath{\langle #1 |}}   
\newcommand{\ket}[1]{\ensuremath{| #1 \rangle}}   
\newcommand{\bfrac}[2]{\left(\frac{#1}{#2}\right)}

\newcommand{\sla}[1]{#1 \hspace*{-1ex}/}

\newcommand{\capdef}{}
\newcommand{\mycaption}[2][\capdef]{\renewcommand{\capdef}{#2}%
       \caption[#1]{{\footnotesize #2}}}
\makeatletter
\renewcommand{\fnum@table}{\textbf{\tablename~\thetable}}
\renewcommand{\fnum@figure}{\textbf{\figurename~\thefigure}}
\makeatother

\newcommand{\wes}{{\ensuremath{\rm WES}}}
\newcommand{\wa}{{\ensuremath{\rm WAS}}}
\newcommand{\wae}{{\ensuremath{\rm el\mbox{-}WAS}}}
\newcommand{\wai}{{\ensuremath{\rm ie\mbox{-}WAS}}}
\newcommand{\wns}{{\ensuremath{\rm WNS}}}

\begin{document}
\pagestyle{plain}

\vspace*{1cm}
\preprint{CERN-PH-TH/2009-116}

\title{DAMA/LIBRA and leptonically interacting Dark Matter\vspace*{1cm}}

\author{\textbf{Joachim Kopp}\vspace*{3mm}}
\email{jkopp AT mpi-hd.mpg.de}
\affiliation{Max-Planck-Institute for Nuclear Physics,
             PO Box 103980, 69029 Heidelberg, Germany}

\author{\textbf{Viviana Niro}\vspace*{3mm}}
\email{niro AT mpi-hd.mpg.de}
\affiliation{Max-Planck-Institute for Nuclear Physics,
             PO Box 103980, 69029 Heidelberg, Germany}

\author{\textbf{Thomas Schwetz}\vspace*{3mm}}
\email{schwetz AT mpi-hd.mpg.de}
\affiliation{Max-Planck-Institute for Nuclear Physics,
             PO Box 103980, 69029 Heidelberg, Germany}

\author{\textbf{Jure Zupan}\vspace*{3mm}}
\email{jure.zupan AT cern.ch}
\affiliation{Theory Division, Physics Department, CERN, CH-1211 Geneva 23, Switzerland}
\altaffiliation{On leave of absence from Faculty of mathematics and physics, University of
  Ljubljana, Jadranska 19, 1000 Ljubljana, Slovenia, and Josef Stefan Institute, Jamova 39, 
  1000 Ljubljana, Slovenia}


\begin{abstract}
  \vspace*{5mm} We consider the hypothesis that Dark Matter (DM) has
   tree-level interactions only with leptons. Such a framework, where DM
   recoils against electrons bound in atoms, has been proposed as an
   explanation for the annually modulated scintillation signal in DAMA/LIBRA
   versus the absence of a signal for nuclear recoils in experiments like
   CDMS or XENON10. 
   However, even in such a leptophilic DM scenario there are loop induced
   DM--hadron interactions, where photons
   emitted from virtual leptons couple to the charge of a nucleus. 
   Using a
   general effective field theory approach we show that, if such an
   interaction is induced at one or two-loop level,
   then 
    DM--nucleus scattering
   dominates over DM--electron scattering. This is because the latter is
   suppressed by the bound state wave function. One obtains a situation
   similar to standard DM--nucleus scattering 
   analyses 
   with considerable tension
   between the results of DAMA and CDMS/XENON10. 
   This conclusion does not
   apply in the case of pseudoscalar or axial vector coupling between DM and leptons, where
   the loop diagrams vanish. In this case the explanation of the DAMA signal
   in terms of DM--electron scattering is strongly disfavored by the
   spectral shape of the signal. Furthermore, if DM can annihilate into
   neutrinos or tau leptons, the required cross sections are excluded by
   many orders of magnitude using the Super-Kamiokande bound on neutrinos
   from DM annihilations in the Sun.

\end{abstract}
\maketitle


\section{Introduction}

The DAMA collaboration provided strong evidence for an annually modulated
signal in the scintillation light from sodium iodine detectors. The combined
data from DAMA/NaI~\cite{Bernabei:2003za} (7 annual cycles) and
DAMA/LIBRA~\cite{Bernabei:2008yi} (4 annual cycles) with a total exposure of
0.82~ton~yr show a modulation signal with $8.2\sigma$ significance. The
phase of this modulation agrees with the assumption that the signal is due
to the scattering of Weakly Interacting Massive Particles (WIMPs) forming
the Dark Matter (DM) halo of our Galaxy.

However, many interpretations of this signal in terms of DM scattering are
in conflict with constraints from other DM direct detection experiments.
Spin-independent elastic WIMP--nucleon scattering accounting for the DAMA
modulation is tightly constrained by bounds from several experiments, most
notably from CDMS~\cite{Ahmed:2008eu} and XENON10~\cite{Angle:2007uj}.
While conventional WIMPs with masses $m_\chi \gtrsim 50$~GeV are excluded by
many orders of magnitude, light WIMPs with $\lesssim 10$~GeV masses might be
marginally compatible with the constraints, see, e.g.~\cite{Bottino:2007qg,
Bottino:2008mf, Petriello:2008jj, Feng:2008dz, Chang:2008xa,
Fairbairn:2008gz, Savage:2008er, Kim:2009ke} for recent studies. 
Spin-dependent couplings to protons can account for the DAMA signal without
being in conflict with CDMS and XENON10, but in this case strong constraints
from COUPP~\cite{Behnke:2008zza}, KIMS~\cite{Lee:2007qn},
PICASSO~\cite{Archambault:2009sm} as well as (somewhat model dependent)
bounds from Super-Kamiokande~\cite{Desai:2004pq} searches for neutrinos from
DM annihilations inside the Sun apply~\cite{Savage:2008er, Hooper:2008cf}.
Inelastic scattering of a DM particle to a nearly degenerate excited state has
been proposed in~\cite{TuckerSmith:2001hy}, see~\cite{Chang:2008gd,
MarchRussell:2008dy, Cui:2009xq, Arina:2009um, SchmidtHoberg:2009gn} for recent
analyses, though also in this case tight constraints apply, in particular from
CRESST-II~\cite{Angloher:2008jj} and ZEPLIN-II~\cite{Cline:2009xd}. Other
proposals include mirror world DM~\cite{Foot:2008nw} or DM with electric or
magnetic dipole moments~\cite{Masso:2009mu}.

In this work we consider the hypothesis that the DM sector has no
direct couplings to quarks, only to leptons, in particular the
electrons.  While electronic events will contribute to the
scintillation light signal in DAMA, most other DM experiments like
CDMS or XENON reject pure electron events by aiming at a (close to)
background free search for nuclear recoils.  DM scattering off
electrons at rest cannot provide enough energy to be seen in a
detector, however, exploiting the tail of the momentum distribution of
electrons bound in an atom may lead to a scintillation light signal in
DAMA of order few keV~\cite{Bernabei:2007gr}. The signal in direct
detection experiments from DM--electron scattering has been considered
recently also in ref.~\cite{Dedes:2009bk}.
An affinity of DM to leptons might also be motivated by recent cosmic
ray anomalies~\cite{Adriani:2008zr, atic, Abdo:2009zk} observed in
electrons/positrons, but not in anti-protons. A simple model for
``leptophilic'' DM has been presented in ref.~\cite{Fox:2008kb}, see
in this context for example also~\cite{Cao:2009yy, Ibarra:2009bm} and
references therein.

In the following we will use effective field theory to
perform a model independent analysis.  We will consider all possible
Lorentz structures for the effective DM--electron interaction and show
that in many cases a DM--quark interaction is induced at
 one or
two-loop level by photon exchange. In these cases the loop induced
DM--nucleon scattering always dominates, since the DM--electron
scattering cross section is suppressed by the momentum wave
function. 
This reintroduces the tension between DAMA
and other searches. 
We identify only one possible Lorentz structure, the axial vector type coupling, 
where  DM--electron scattering dominates and the scattering cross section 
is not additionally suppressed by small quantities. 
Taking special
care of the kinematics in the DM scattering off bound electrons, we
show that in this case the fit to the DAMA event spectrum is very
bad. Super-Kamiokande constraints on neutrinos from DM annihilations
inside the Sun 
also disfavor this possibility. 
Our results thus
suggest that leptonically interacting DM is not a viable explanation
of the DAMA annual modulation signal.

The plan of the paper is as follows. In sec.~\ref{sec:II} we introduce the
effective Lagrangian for DM--lepton interactions, discuss three possible
experimental signatures of leptophilic DM, and estimate their relative
sizes. In sec.~\ref{sec:lorentz} we discuss all the possible Lorentz
structures and the implications for tree-level interactions with electrons
as well as loop induced DM--nucleus interactions. In sec.~\ref{sec:rates}
the event rates in direct detection experiments are calculated, while
sec.~\ref{sec:results} contains the numerical results for two representative
examples, namely vector like couplings, where the count rate is
dominated by loop induced WIMP--nucleus scattering
(sec.~\ref{sec:results-V}), and axial vector coupling, where no loop
contribution is present and WIMP--electron scattering off bound electrons
dominates (sec.~\ref{sec:results-A}). In sec.~\ref{sec:SK} we show that the
cross sections required in the axial vector case are ruled out by
Super-Kamiokande constraints assuming that DM annihilations in the Sun
produce neutrinos; we point out the importance of the non-zero temperature
of the electrons in the Sun. We summarize our results in
sec.~\ref{sec:concl}. Technical details and supplementary information is
given in appendices~\ref{app:loop}, \ref{app:WE}, \ref{app:wave-function},
\ref{app:equil}.

\section{Leptonically interacting dark matter}
\label{sec:II}

\subsection{Effective dark matter interactions}
\label{sec:vertex}

\begin{figure}
  \begin{center}
    \includegraphics[width=15cm]{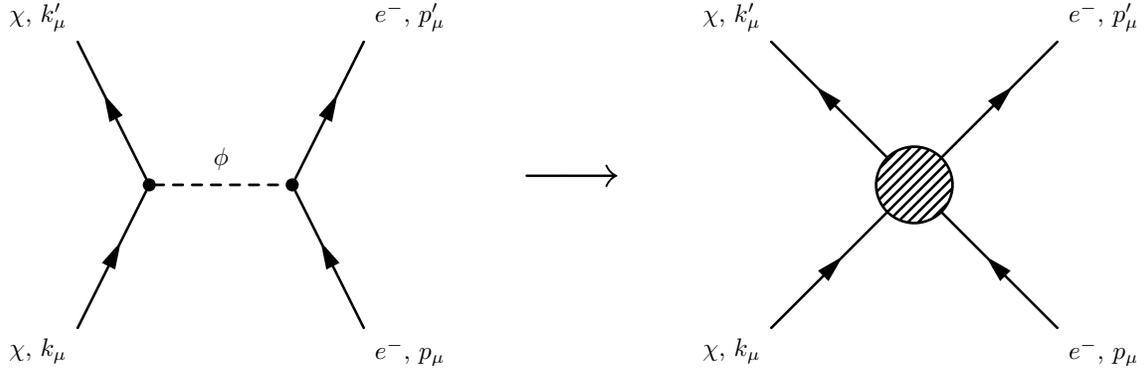}
  \end{center}
  \caption{Example for generating an effective local DM-electron interaction
  vertex (right diagram) as used in our analysis by the exchange of a heavy
  intermediate particle $\phi$ (left diagram).} \label{fig:4f-diagram}
\end{figure}

The goal of our study is a model independent analysis under the assumption
that the DM particle $\chi$ couples directly only to leptons but not to
quarks. The appropriate description is in terms of effective interactions. 
Let us first focus on the case of fermionic DM. The most general dimension
six four-Fermi effective interactions are then, shown pictorially also in
fig.~\ref{fig:4f-diagram} (right diagram),
\be\label{eq:4fermi}
{\cal L}_{\rm eff}=\sum_i G \, (\bar\chi \Gamma_\chi^i \chi) \, (\bar\ell \Gamma_\ell^i \ell) 
\qquad\text{with}\qquad
G = \frac{1}{\Lambda^2}\,,
\ee
where $\Lambda$ is the cut-off scale for the effective field theory
description, while the sum is over different Lorentz structures. A complete
set consists of scalar ($S$), pseudo-scalar ($P$), vector ($V$),
axial-vector ($A$), tensor ($T$), and axial-tensor ($AT$) currents. The
four-Fermi operators can thus be classified to be of
\be\label{eq:types}
\begin{array}{l@{\qquad}l@{\qquad}l}
  \text{scalar-type:} & 
  \Gamma_\chi = c^\chi_S + ic^\chi_P\gamma_5, & 
  \Gamma_\ell = c^\ell_S + ic^\ell_P\gamma_5, \\
  \text{vector-type:} & 
  \Gamma_\chi^\mu = (c^\chi_V + c^\chi_A\gamma_5)\gamma^\mu, & 
  \Gamma_{\ell\mu} = (c^\ell_V + c^\ell_A\gamma_5)\gamma_\mu ,\\
  \text{tensor-type:} &
  \Gamma_\chi^{\mu\nu} = (c_T + ic_{AT} \gamma_5)\sigma^{\mu\nu}, &
  \Gamma_{\ell\mu\nu} = \sigma_{\mu\nu}, 
\end{array}
\ee
where $\sigma_{\mu\nu} = \frac{i}{2}[\gamma_\mu,\gamma_\nu]$.\footnote{The
relation $\sigma^{\mu\nu}\gamma_5 = \frac{i}{2}
\epsilon^{\mu\nu\alpha\beta}\sigma_{\alpha\beta}$ implies that the
$AT\otimes AT$ coupling is equivalent to $T\otimes T$, and $T\otimes AT = AT
\otimes T$.} If DM is a Majorana particle, vector and tensor like
interactions are forbidden, i.e., $c^\chi_V = c^\chi_T = c^\chi_{AT} = 0$.

In our work we do not rely on any specific realization of the effective
interaction. The simplest example would just be assuming that the
interaction is induced by the exchange of an intermediate particle whose
mass is much larger than the recoil momenta that are of order a few MeV. 
The intermediate particle can then be integrated out leaving an effective
point interaction. Let us look at the $\chi$-lepton interaction mediated by
a scalar field $\phi$, shown in fig.~\ref{fig:4f-diagram}. It gives an
amplitude 
\be i g_S^\chi (\bar u_\chi
u_\chi) \; \frac{i}{q^2-m_\phi^2+i\epsilon}\;i g_S^\ell (\bar u_\ell
u_\ell)
\quad\longrightarrow\quad
i \; \frac{g_S^\chi g_S^\ell}{m_\phi^2}\;(\bar u_\chi u_\chi)
(\bar u_\ell u_\ell) \,,
\ee 
where on the right-hand side we have neglected the momentum transfer
$q^2=(p'-p)^2\ll m_\phi^2$. The same amplitude is obtained from a local
operator $(\bar\chi \chi) \, (\bar\ell \ell)$ with a Wilson coefficient
${g_S^\chi g_S^\ell}/{m_\phi^2}$ (in the notation used in
eqs.~\ref{eq:4fermi}, \ref{eq:types} we have $c_S^\chi=g_S^\chi,
c_S^\ell=g_S^\ell, \Lambda=m_\phi$).

In the case of scalar DM, at lowest order there is only one
dimension five operator. The effective Lagrangian is given by 
\be
{\cal L}_{\rm eff}=G_5 (\chi^\dagger \chi)\left[\bar\ell(d_S + id_P\gamma_5)\ell\right]
\qquad\text{with}\qquad
G_5 = \frac{1}{\Lambda}\,.\label{scalar:G5}
\ee

\subsection{Signals in direct detection experiments}
\label{sec:signals}

\begin{figure}
  \begin{center}
    \includegraphics[height=4.3cm]{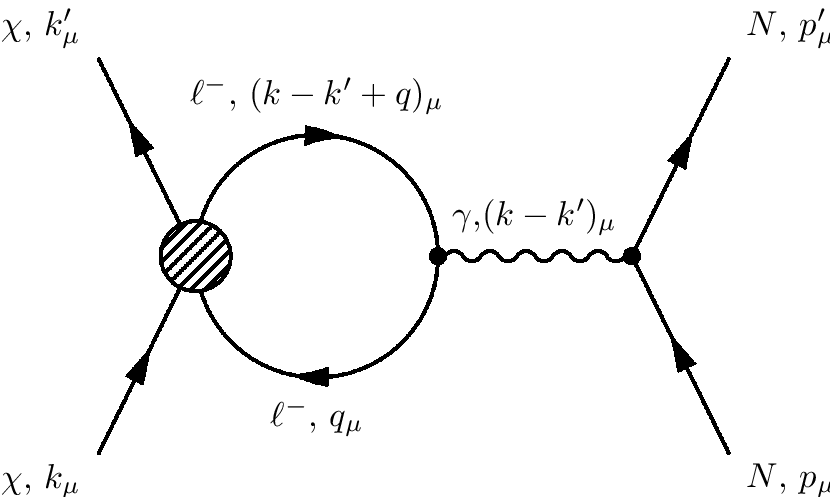}\\[6mm]
    \includegraphics[height=3.3cm]{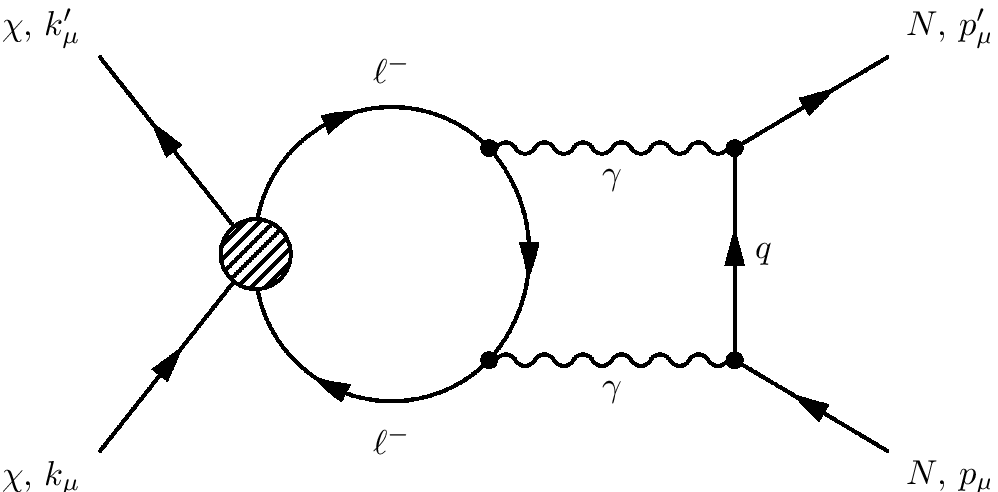}\qquad\qquad
    \includegraphics[height=3.3cm]{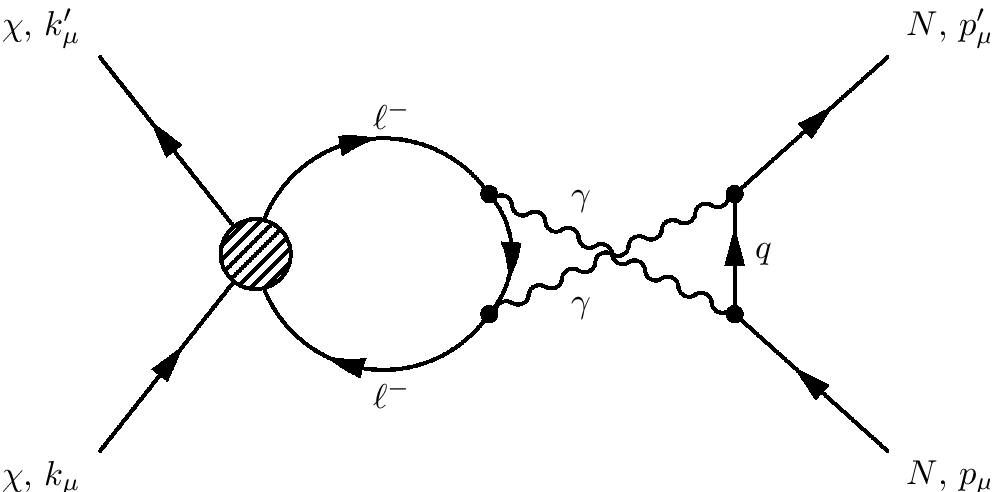}
  \end{center}
  \caption{DM--nucleus interaction induced by a charged lepton loop and photon
  exchange at 1-loop (top) and 2-loop (bottom).}
  \label{fig:loop-diagram}
\end{figure}

Let us now discuss the signals that arise when a ``leptophilic'' DM particle
interacts in a detector. One can distinguish the following types of signals
(see also~\cite{Dedes:2009bk}):
\begin{enumerate}
  \item {\it WIMP--electron scattering} (\wes): The whole recoil is
    absorbed by the electron that is then kicked out of the atom to which it
    was bound.

  \item {\it WIMP--atom scattering} (\wa): The electron on which the DM
    particle scatters remains bound and the recoil is taken up by the whole
    atom. The process can either be elastic (\wae) in which case the
    electron wave function remains the same, or inelastic (\wai), in which
    case the electron is excited to an outer shell.
    
  \item Loop induced {\it WIMP--nucleus scattering} (\wns): Although per assumption DM
    couples only to leptons at tree level, an
    interaction with quarks is induced at loop level, by coupling a photon
    to virtual leptons, see fig.~\ref{fig:loop-diagram}. This will lead to
    scattering of the DM particle off nuclei.  
\end{enumerate}

\wes\ produces a prompt electron and possibly additional Auger electrons or
X-rays. This leads to a signal in scintillation detectors such as DAMA, but
is rejected in nuclear recoil experiments like CDMS and XENON. If \wes\
was the dominant mechanism, it might be possible to explain both the DAMA
signal and the absence of the signal in CDMS and XENON. In the other two
cases, \wa\ and \wns, the signal consists of a scattered nucleus and shows
up in all direct detection experiments searching for DM nuclear recoils. If
\wa\ or \wns\ was the dominant signal, then the leptophilic nature of DM
would not help to resolve the tension between DAMA and the remaining
experiments. In the following we will first give rough estimates for the
relative sizes of these three signal types for different DM--lepton effective
interactions, while giving a more detailed calculation of the event rates
later in sec.~\ref{sec:rates} and appendix~\ref{app:WE}.

The event rate in direct detection experiments is proportional to the
differential cross section $d\sigma/d E_d$, where 
\be
E_d = E_\chi - E'_\chi \,,
\label{eq:Econs}
\ee
is the energy deposited by the WIMP in the detector. The DAMA annual
modulation signal is observed at $E_d \simeq 3$~keV. Also for other
direct detection experiments typical values are in the few to tens of keV range.
As we will see in sec.~\ref{sec:rates} and appendix~\ref{app:WE}, just from
kinematics the cross section is proportional to
\be\label{eq:scaling}
\frac{d\sigma}{dE_d} \propto G^2 m_e \, (G^2 m_N) 
\quad\text{for}\quad
\wes\,(\wa, \wns)  \,,
\ee
where $G$ is defined in eq.~\ref{eq:4fermi} and $m_e$ ($m_N$) is the
electron (nucleus) mass. This suppresses the \wes\ induced event rate by a
factor $m_e/m_N$ with respect to \wa\ and \wns.

\begin{figure}
  \begin{center}
    \includegraphics[width=\textwidth]{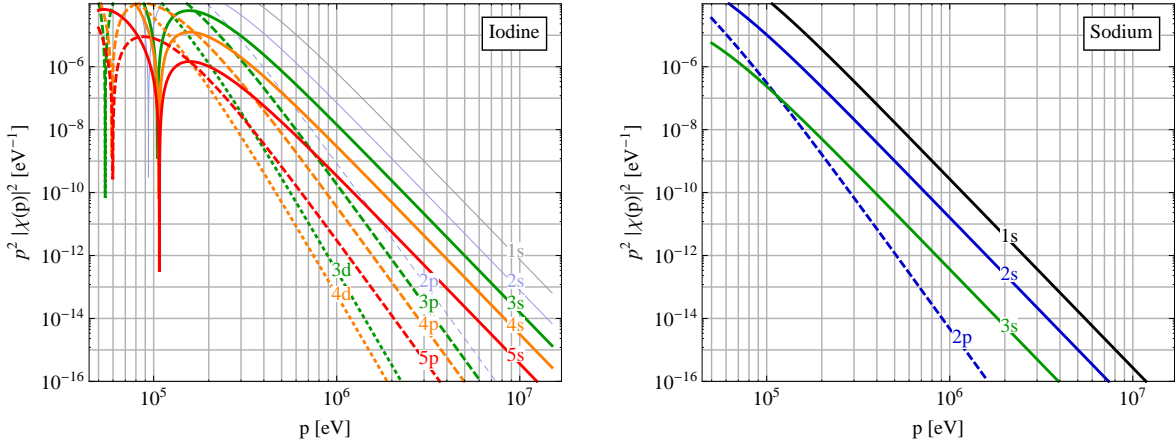}
  \end{center}
  \vspace{-1.5em}
  \caption{Momentum space wave functions of iodine and sodium. Solid colored curves
    correspond to shells that contribute to \wes\ in DAMA, while thin
    light curves are for shells that are not accessible in DAMA.}
  \label{fig:wf}
\end{figure}

In order for \wes\ to deposit $\sim$~keV energy in the detector, the
electron that a WIMP scatters off has to have quite a high momentum.
Indeed, the maximal detectable energy from DM scattering on electrons at
rest is $2 m_e v^2$, with typical DM velocities of $v\sim 10^{-3}c$. Hence,
the maximal detectable energy is of order eV, far too low to be relevant for
the DAMA signal at few keV. Therefore, one has to explore the scattering off
bound electrons with non-negligible momentum~\cite{Bernabei:2007gr}.  In
this case the energy transfer to the detector is $E_d\sim \mathcal{O}(p v)$,
and an electron momentum $p \sim$~MeV is required to obtain $E_d \sim$~keV.
Since electrons are bound in the atom, there is a finite yet small
probability that it carries such high momentum. The detailed calculations
below will show that the suppression factor from the wave function is given
by the expression
\be\label{eq:wf-sup}
  \epsilon_\wes =
    \sqrt{2 m_e (E_d - E_{B})} \, (2l+1)
    \int\!\frac{dp\,p}{(2\pi)^3} \,  |\chi_{nl}(p)|^2  \sim 10^{-6}\,.
\ee
The integral is over MeV momenta, while $\chi_{nl}(p)$ is the momentum
wave function of the shell $nl$ with the binding energy $E_B$. Some wave
functions are shown in fig.~\ref{fig:wf}, which we have used to obtain the
numerical estimate for $\epsilon_\wes \sim 10^{-6}$ given above. 

Similarly, \wa\ is also suppressed by the overlap of atomic wave functions of
the initial and final states of the electron:
\be\label{eq:wf-was}
\epsilon_\wa=
\sum
|\bra{n'l'm'} e^{i (\vect{k} - \vect{k'}) \vect{x}}\ket{nlm}|^2 
\sim 10^{-19} \,,
\ee
where the numerical estimate follows from fig.~\ref{fig:me-wai} in
appendix~\ref{app:wave-function} for a momentum transfer of
$|\vect{k}-\vect{k}'| = \sqrt{2 m_N(E_d -\delta E_B)} \sim 10$~MeV.

Loop induced \wns\ does not suffer from any wave function suppression, but
instead 
carries a loop factor. At 1-loop the suppression is of order
$(\alpha_\mathrm{em} Z / \pi)^2$, with $Z$ being the charge number of the
nucleus. Combining this with eqs.~\ref{eq:scaling}, \ref{eq:wf-sup},
\ref{eq:wf-was}, we obtain the following rough estimate for the ratios of
\wa, \wes\ and \wns\ induced event rates 
(neglecting order-one factors but also 
possible different $v$
dependences):
\be\label{eq:hierarchy}
R^\wa:R^\wes:R^\wns \sim
\epsilon_\wa : 
\epsilon_\wes \, \frac{m_e}{m_N}: 
\left(\frac{\alpha_{\rm em}Z}{\pi}\right)^2 
 \sim 10^{-17}:10^{-10}:1 \, ,
\ee
where in the last step we used $m_N=100$~GeV and $Z=53$. 

We conclude that whenever a loop induced cross section is present it will
dominate the rate in direct detection experiments. This holds for 1-loop as
well as 2-loop cross sections, since the latter will be suppressed by
another factor $(\alpha_\mathrm{em} Z / \pi)^2 \simeq 5\times 10^{-6} Z^2$
relative to 1-loop, and hence they are still much larger than the \wes\
contribution. As we will show in the next section for axial vector like
DM--lepton coupling no loop will be induced. Therefore, in this case the
signal in DAMA will be dominated by \wes. Then, \wa\ is still
irrelevant for DAMA, but since \wes\ will not contribute to the rate in CDMS
and XENON, \wa\ might in principle lead to a signal in those experiments.

\section{DM--electron scattering versus loop induced nuclear recoil}
\label{sec:lorentz}

Having estimated the strong hierarchy among the \wa, \wes, and \wns\ signals
in the previous section, we now discuss which type of signal is present for
a given Lorentz structure of the effective DM--lepton vertex,
eqs.~\ref{eq:types}, \ref{scalar:G5}.

\subsection{DM scattering on electrons}

Let us start by investigating DM scattering on electrons, relevant for \wes\
and \wa. To simplify the discussion we consider $\chi$ scattering on
electrons at rest. 
This will enable us to see for which types of Lorentz
structures in the effective DM-lepton Lagrangian, eq. \ref{eq:4fermi},
this interaction is relevant. We defer the complications
introduced by the fact that electrons are actually bound in atoms to
sec.~\ref{sec:rates} and appendix~\ref{app:WE}.

We consider a DM particle $\chi$ of mass $m_\chi$ scattering elastically on
a free electron at rest, assuming that all the particles are
non-relativistic.  
The scattering cross sections for
fermionic DM are then:
\begin{eqnarray}
\text{scalar-type:} && \sigma = 
\sigma_{\chi e}^0
\left\{
(c^\chi_S c^e_S)^2 +
\left[ (c^\chi_S c^e_P)^2 + (c^\chi_P c^e_S)^2 \frac{m_e^2}{m_\chi^2} \right]
\frac{v^2}{2} + 
\frac{(c^\chi_P c^e_P)^2}{3}\frac{m_e^2}{m_\chi^2} v^4 \right\},
\label{eq:sigS}\\
\text{vector-type:} &&
\sigma = \sigma_{\chi e}^0 
\left\{ (c^\chi_V c^e_V)^2 + 3(c^\chi_A c^e_A)^2
+\left[ (c^\chi_V c^e_A)^2 + 3(c^\chi_A c^e_V)^2 \right]
\frac{v^2}{2}\right\} ,
\label{eq:sigV}\\
\text{tensor-type:} &&
\sigma = \sigma_{\chi e}^0 
\left\{ 12 c_T^2 + 6 c_{AT}^2 v^2 \right\}.
\label{eq:sigT}
\end{eqnarray} 
In the above expressions there are two suppression factors, the DM velocity
in our halo $v\sim 10^{-3}c$ and the ratio $m_e/m_\chi$. The cross section for
each Lorentz structure is given to leading order in these expansion
parameters. Up to the velocity or electron mass suppression the typical
size of the scattering cross section is 
\be\label{eq:sig0ferm}
\sigma_{\chi e}^0 \equiv \frac{G^2 m_e^2}{\pi} 
= \frac{ m_e^2}{\pi \Lambda^4} \approx
3.1\times 10^{-39} \unit{cm}^2 \,  \bfrac{\Lambda}{10 \unit{GeV}}^{-4} \,.
\ee

For scalar DM the $\chi e$ scattering cross section is induced by
the dimension 5 operator, eq.~\ref{scalar:G5}, giving
\be\label{eq:sig_scalarDM}
\sigma = \sigma_{\chi e,5}^0 \left( d_S^2 +  \frac{d_P^2}{2} v^2 \right), 
\ee
with
\be\label{eq:sig0scalar}
\sigma_{\chi e,5}^0 \equiv
\frac{G^2_5}{4\pi} \frac{m_e^2}{m_\chi^2} =
\frac{1 }{4\pi \Lambda^2} \frac{m_e^2}{m_\chi^2} =
7.7\times 10^{-42} \unit{cm}^2 \,  \bfrac{\Lambda}{10 \unit{GeV}}^{-2}
\bfrac{m_\chi}{100 \unit{GeV}}^{-2} \,.
\ee
Compared to fermionic DM two powers of $\Lambda$ are replaced by $m_\chi$
which typically is larger than $\Lambda$. The scalar DM scattering
cross section is thus further suppressed compared to the fermionic case for
given $\Lambda$. The results of eqs.~\ref{eq:sigS}--\ref{eq:sigT} and
\ref{eq:sig_scalarDM} are summarized in the middle column of
Tab.~\ref{tab:Xsec}.

\begin{table}
\begin{tabular}{c@{\quad}|@{\quad}c@{\qquad}cc}
\hline\hline
& \multicolumn{3}{c}{fermionic DM} \\
\hline
$\Gamma_\chi \otimes \Gamma_\ell$ &  $\sigma(\chi e\to \chi e)/\sigma_{\chi
e}^0$& 
\multicolumn{2}{c}{$\sigma(\chi N \to \chi N)/\sigma_{\chi N}^1$} \\
\hline
$S\otimes S$ & $1$ &  $\alpha_{\rm em}^2$ & [2-loop]\\
$S\otimes P$ & $\mathcal{O}(v^2)$        &  $-$ & \\
$P\otimes S$ & $\mathcal{O}(r_e^2 v^2)$          &  $\alpha_{\rm em}^2 v^2$ & [2-loop]  \\
$P\otimes P$ & $\mathcal{O}(r_e^2 v^4)$        & $-$ &   \\
$V\otimes V$ & $1$ &  $1$ & [1-loop]\\
$V\otimes A$ & $\mathcal{O}(v^2)$          & $-$ &  \\
$A\otimes V$ & $\mathcal{O}(v^2)$          &   $v^2$ &[1-loop] \\
$A\otimes A$ & $3$ & $-$ &   \\
$T\otimes T$ & $12$ &  $q_\ell^2$& [1-loop]\\
$AT\otimes T$& $\mathcal{O}(v^2)$          &  $q_\ell^2 v^{-2}$ &[1-loop]\\
\hline\hline
 & \multicolumn{3}{c}{scalar DM} \\
 \hline
$\Gamma_\ell$&  $\sigma(\chi e\to \chi e)/\sigma_{\chi e,5}^0$& 
\multicolumn{2}{c}{$\sigma(\chi N \to \chi N)/\sigma_{\chi N,5}^1$} \\
\hline
$S$  & $1$ & $\alpha_{\rm em}^2$ & [2-loop] \\
$P$  & $\mathcal{O}(v^2)$          & $-$ & \\
\hline\hline
\end{tabular}
  \mycaption{Scattering cross section suppression by small parameters for
  DM--electron scattering and loop induced DM--nucleon scattering for all
  possible Lorentz structures. Here, $v\sim 10^{-3}$ is the DM velocity, 
  $r_e=m_e/m_\chi$, and $q_\ell = m_\ell/m_N$ ($\ell = e,\mu,\tau$). The
  reference cross sections $\sigma_{\chi e}^0$, $\sigma_{\chi e,5}^0$,
  $\sigma_{\chi N}^1$, $\sigma_{\chi N,5}^1$ are defined in
  eqs.~\ref{eq:sig0ferm}, \ref{eq:sig0scalar}, \ref{eq:sigma1}. The
  couplings $c^\chi, c^\ell, d$ have been set to one. The entries for $\chi
  N\to \chi N$ are orders of magnitude estimates.  \label{tab:Xsec}}
\end{table}

\subsection{Loop induced DM--nucleus interactions}
\label{sec:loop}

We have assumed that DM is leptophilic, so that at scale $\Lambda$ only
operators connecting DM to leptons, eqs. \ref{eq:4fermi}, \ref{eq:types},
\ref{scalar:G5}, are generated. However, even under this assumption, at loop
level one does induce {\it model independently} also couplings to quarks
from photon exchange between virtual leptons and the quarks.  The diagrams
that can arise at one and two-loop order are shown in
fig.~\ref{fig:loop-diagram}.\footnote{Similar diagrams with a photon
replaced by a $Z^0$ or a Higgs boson are power suppressed by
$(k-k')^2/M_{Z^0,H}^2$ and thus negligible.} The lepton running in the loop
can be either an electron or any other charged lepton to which the DM
couples.

The one loop contribution involves the integral over loop momenta of the form
\be
\int \frac{d^4q}{(4\pi)^4}\mathrm{Tr}\left[
\Gamma_\ell \, \frac{\sla{q}' + m_\ell}{{q'}^2 -m_\ell^2} \,
\gamma^\mu \, \frac{\sla{q} + m_\ell}{q^2 - m_\ell^2} \right],
\ee
with $q'=k-k'+q$ and $k,k'$ the incoming momenta as denoted in
fig.~\ref{fig:loop-diagram} and $\Gamma_\ell$ the Dirac structures given in
eqs. \ref{eq:types}, \ref{scalar:G5}. The one loop contribution is nonzero
only for vector and tensor lepton currents,
$\Gamma_\ell=\gamma_\mu,\sigma_{\mu\nu}$. For the scalar lepton current,
$\Gamma_\ell = 1$, the loop integral vanishes, reflecting the fact that one
cannot couple a scalar current to a vector current. 
The DM--quark interaction is then induced at two-loops through the diagrams
shown in fig.~\ref{fig:loop-diagram}. In contrast for pseudo-scalar and
axial vector lepton currents, $\Gamma_\ell = \gamma_5, \gamma_\mu \gamma_5
$, the diagrams vanish to all loop orders. One insertion of $\gamma_5$ gives
either zero or a fully anti-symmetric tensor $\epsilon^{\alpha\beta\nu\mu}$.
Since there are only three independent momenta in a $2 \to 2$ process, two
indices need to be contracted with the same momentum, yielding zero. 

The calculation of the 1-loop and 2-loop cross sections for scattering of DM
on a nucleus is relegated to appendix~\ref{app:loop}. There we give the full
1-loop expressions, whereas here we collect the main results in the ``leading log'' approximation, 
neglecting the remaining logarithmic dependence on momentum transfer. 
The approximate 2-loop results are obtained in the limit of heavy leptons. Expanding also 
 in the $\chi$
velocity $v\sim 10^{-3}$ to first non-zero order, the differential cross sections
$d\sigma/dE_d$ are 
\begin{align}
\text{vector type:}~~~~\frac{d\sigma}{dE_d}=&\frac{d\sigma_N^1}{dE_d} 
\Big[\log\Big(\frac{m_\ell^2}{\mu^2}\Big)\Big]^2 \frac{1}{9}
\Big\{(c_V^\chi c_V^\ell)^2+(c_A^\chi c_V^\ell)^2\Big[v^2+v_d^2\Big(2-\frac{m_N^2}{\mu_N^2}\Big)\Big]\Big\}
F(q)^2 \,,\label{eq:vector:nucl}\\
\text{tensor type:}~~~~\frac{d\sigma}{dE_d}=&\frac{d\sigma_N^1}{dE_d} 
\Big[\log\Big(\frac{m_\ell^2}{\mu^2}\Big)\Big]^2 \frac{4}{v_d^2}\frac{m_\ell^2}{m_N^2}
\big\{c_T^2 v^2+c_{AT}^2\big\} F(q)^2 \,,\label{eq:tensor:nucl}\\
\text{scalar type:}~~~~\frac{d\sigma}{dE_d}=&\Big(\frac{\alpha_{\rm em}Z}{\pi}\Big)^2
\frac{d\sigma_N^1}{dE_d} \Big(\frac{\pi^2}{12}\Big)^2 \frac{m_N^2}{m_\ell^2}v_d^2
\Big\{(c_S^\chi c_S^\ell)^2+\frac{1}{4}(c_P^\chi c_S^\ell)^2v_d^2 \frac{m_N^2}{m_\chi^2}\Big\}
\tilde F(q)^2 \,,\label{scalar:two-loop:diff}\\
\text{scalar DM:}~~~~\frac{d\sigma}{dE_d}=&\Big(\frac{\alpha_{\rm em}Z}{\pi}\Big)^2
\frac{d\sigma_{N,5}^1}{dE_d} \Big(\frac{\pi^2}{12}\Big)^2 \frac{m_N^2}{m_\ell^2}v_d^2
(d_S^\ell)^2 \tilde F(q) \,,\label{scalarDM:two-loop:diff}
\end{align}
where the common 1-loop cross section prefactor is
\be\label{diff:sigma1} 
  \frac{d\sigma_N^1}{dE_d} = \frac{m_N}{2\pi\, v^2}
  \Big(\frac{\alpha_{\rm em}Z}{\pi}G\Big)^2 \,, 
\ee 
and $d\sigma_{N,5}^1/dE_d$ is given by the same expression with $G\to
G_5/(2m_\chi)$. Here $m_N$ and $Z$ are the nucleus mass and charge,
respectively, while $\mu_N = m_N m_\chi/(m_N + m_\chi)$ is the reduced mass
of the two-particle system. The two small parameters are the $\chi$ velocity
$v\sim 10^{-3}$ and the velocity of the recoiled nucleus, $v_d=\sqrt{2
E_d/m_N}$. The kinetic recoil energy of the nucleus $E_d$ in the $\chi N\to \chi N$
scattering, cf. eq.~\ref{eq:Econs}, 
has a size $E_d \sim$~keV.

In the calculations we set $\mu=\Lambda$, since this is the scale at which 
the Wilson coefficient $G$ is generated. The form factors $F(q)$ and $\tilde
F(q)$ account for the nuclear structure. For the form factor $F(q)$ entering the one 
loop induced scattering cross section we
use~\cite{Jungman:1995df} $F(q) = 3 e^{-\kappa^2 s^2/2} [\sin(\kappa
r)-\kappa r\cos(\kappa r)] / (\kappa r)^3$, with $s = 1$~fm, $r = \sqrt{R^2
- 5 s^2}$, $R = 1.2 A^{1/3}$~fm, $\kappa = \sqrt{2 m_N E_d}$ (and
$q^2\simeq-\kappa^2$).
 The form factor $\tilde F(q)$ entering the 2-loop
expressions accounts for nuclear structure in the case of two-photon exchange. 
Its precise form is not needed in the subsequent analysis, though.
The two-loop scalar type differential cross section in
eq.~\ref{scalar:two-loop:diff} was calculated integrating out first the leptons assuming they are heavy.
This is an appropriate limit for muon and tau intermediate states, where
$m_\mu, m_\tau\gg \kappa$, while for electrons $m_e\sim \kappa$ and the
expression for the cross section is only approximate, see
appendix~\ref{app:loop} for details.

For easier comparison with the previous subsection we also quote the results
for the total $\chi N\to \chi N$ cross sections, integrated over the recoil
energy $E_d$. For simplicity we neglect the dependence on the nuclear form
factors and set $F(q)=\tilde F(q)=1$ for this comparison, giving
\begin{align}
\text{vector type:}~~~~\sigma=&\sigma_N^1 \Big[\log\Big(\frac{m_\ell^2}{\mu^2}\Big)\Big]^2\frac{1}{9}\Big\{(c_V^\chi c_V^\ell)^2+(c_A^\chi c_V^\ell)^2v^2\Big[1+\frac{1}{2}\frac{\mu_N^2}{m_N^2}\Big]\Big\},\label{vec:full}\\
\text{tensor type:}~~~~\sigma=&\sigma_N^1 \Big[\log\Big(\frac{m_\ell^2}{\mu^2}\Big)\Big]^2\frac{m_\ell^2}{\mu_N^2}\Big\{c_T^2 +c_{AT}^2\frac{1}{v^2}\Big\}\log\Big(\frac{E_d^{\rm max}}{E_d^{\rm min}}\Big),\\
\text{scalar type:}~~~~\sigma=&\Big(\frac{\alpha_{\rm em}Z}{\pi}\Big)^2\sigma_N^1 \Big(\frac{\pi^2}{12}\Big)^2\Big(\frac{\mu_N v}{m_\ell}\Big)^2\Big\{2 (c_S^\chi c_S^\ell)^2 +\frac{4}{3}(c_P^\chi c_S^\ell)^2v^2 \frac{\mu_N^2}{m_\chi^2}\Big\},\\
\text{scalar DM:}~~~~\sigma=&\Big(\frac{\alpha_{\rm em}Z}{\pi}\Big)^2\sigma_{N,5}^1 \Big(\frac{\pi^2}{12}\Big)^2\Big(\frac{\mu_N v}{m_\ell}\Big)^2 2 (d_S^\ell)^2,
\end{align}
where $\sigma_N^1$ is the integral of the differential cross section of eq.~\ref{diff:sigma1}
\be\label{eq:sigma1}
\sigma_N^1=\frac{\mu_N^2}{\pi}\Big(\frac{\alpha_{\rm em}Z}{\pi}G\Big)^2\approx
1.9 \times 10^{-32} \unit{cm}^2 \,  \bfrac{\Lambda}{10 \unit{GeV}}^{-4} \bfrac{\mu_N}{10 \unit{GeV}}^{2}  \bfrac{Z}{53}^{2}\,,
\ee
and $\sigma_{N,5}^1$ is obtained from the above expression with $G\to G_5/(2m_\chi)$.
The above results are summarized in table~\ref{tab:Xsec}, facilitating
comparison with $\chi$ scattering on free electrons. In table~\ref{tab:Xsec}
we took $\mu_N\sim m_N\sim m_\chi$, while the scaling for other values of
nucleon and DM masses is easy to obtain from above results.

\subsection{Discussion of Lorentz structures}
\label{sec:ref}

In sec.~\ref{sec:signals} we have estimated a strong hierarchy between the
three types of signals as $R^\wa \ll R^\wes \ll R^\wns$, see
eq.~\ref{eq:hierarchy}. These results imply that whenever \wns\ at 1-loop or
2-loop is generated it 
dominates the event rate in direct detection
experiments. The Lorentz
structures for which this situation applies can be read off from table~\ref{tab:Xsec}. To be specific we will use 
as a representative example of this class the $V\otimes V$ coupling in
the rest of this paper. From the table we also see that there is one case --- the
$A\otimes A$ coupling --- where no $\chi N$ scattering is induced at loop level and moreover the WIMP--electron cross section is not additionally $v$ and/or $m_e/m_\chi$ suppressed. Hence, we chose the
$A\otimes A$ coupling as our second representative example to 
quantitatively discuss the case of a \wes\ dominated event rate. The
results from these two examples can be qualitatively extrapolated to other Lorentz
structures using table~\ref{tab:Xsec}. 

As we will see in the following, the $\chi e\to \chi e$ cross section in the
$A\otimes A$ case has to be very large (corresponding to $\Lambda\sim \mathcal{O}(100 {\text{~MeV}}))$ in order to be relevant for DAMA. 
For the cases in table~\ref{tab:Xsec} where
$\sigma_{\chi e}^0$ is further suppressed by small numbers, like for example
$S\otimes P$ or $P\otimes P$ the scale $\Lambda$ would have to be even lower, so that the effective field theory
 description would break down.

Finally, let us mention the tensor coupling $T\otimes T$, where the
1-loop cross section is suppressed by $m_\ell^2/m_N^2$, while $\chi e$
scattering is enhanced by a factor 12. If DM couples {\it only} to the
electron and not to $\mu$ and $\tau$ the suppression of the loop is of order
$m_e^2/m_N^2 \sim 10^{-10}$, and hence, \wes\ and \wns\ rates can be of
comparable size. However, in general one expects also a coupling to the
$\mu$ and $\tau$ leptons. To be specific, in our numerical analysis of $V\otimes V$ and $A\otimes A$
cases we will assume equal couplings to all three leptons. For the tensor case the same choice would mean
that \wns\ dominates.

\section{Event rates}
\label{sec:rates}

In this section we provide the event rates in direct detection experiments.
For \wes\ and \wa\ we assume $A\otimes A$ coupling and for \wns\ we take
$V\otimes V$. These rates will be used for the numerical fits to DAMA, CDMS,
and XENON data in the following. As argued above, the $A\otimes A$ and
$V\otimes V$ cases are representative 
enough to cover qualitatively all possible
Lorentz structures.

The differential counting rate in a direct DM detection experiment (in units
of counts per energy per kg detector mass per day) is given by
\begin{align}
  \frac{dR}{dE_d} = \frac{\rho_0}{m_\chi} \frac{\eta}{\rho_\mathrm{det}} \,
    \int\!d^3v \, \frac{d\sigma}{dE_d} \, v f_\odot(\vect{v}) \, ,
  \label{eq:dRdE-1}
\end{align}
where $E_d = E_\chi - E'_\chi$ is the energy deposited in the detector,
$\rho_0$ is the local DM density (which we take to be $0.3\
\text{GeV}\,\text{cm}^{-3})$, $\eta$ is the number density of target
particles, and $\rho_\mathrm{det}$ is the mass density of the detector. If
the target contains different elements (like in the case of the DAMA NaI
crystals), the sum over the corresponding counting rates is implied. 

In eq.~\ref{eq:dRdE-1}, $f_\odot(\vect{v})$ is the local WIMP velocity
distribution in the rest frame of the detector, normalized according to
$\int d^3v \, f_\odot(\vect{v}) = 1$. It follows from the DM velocity
distribution in the rest frame of the galaxy, $f_\mathrm{gal}(\vect{v})$, by
a Galilean transformation with the velocity of the Sun in the galaxy and
the motion of the Earth around the Sun. For $f_\mathrm{gal}(\vect{v}) =
f_\mathrm{gal}(v)$ we assume the conventional Maxwellian distribution with
$\bar v = 220\,\rm km\, s^{-1}$ and a cut-off due to the escape velocity
from the galaxy of $v_{\rm esc} = 650\,\rm km\, s^{-1}$: $f_\mathrm{gal}(v)
\propto \exp(-v^2/\bar{v}^2) - \exp(v_{\rm esc}^2/\bar{v}^2)$ for $v \leq
v_{\rm esc}$ and zero for $v > v_{\rm esc}$. We have checked that the
precise value of the escape velocity has a negligible impact on our results.

A scattered nucleus does not deposit all its energy in 
the  form of scintillation
light. This effect is taken into account by the so-called quenching factors,
which are $q_{\rm Na} \simeq 0.3$ for sodium recoils and $q_{\rm I} \simeq
0.085$ for iodine recoils~\cite{Bernabei:1996vj} in DAMA.  In
refs.~\cite{Drobyshevski:2007zj, Bernabei:2007hw} it has been pointed out
that the so-called channeling effect could be relevant, implying that for a
certain fraction of events no quenching would occur due to the special
orientation of the recoil with respect to the crystal. So far this effect
has not been confirmed experimentally in the relevant energy range. Indeed,
the results of ref.~\cite{Graichen:2002kg} do not indicate the presence of
any variation of the count rate for special crystal directions.  Quenching
and channeling is relevant in DAMA in the cases of \wa\ and \wns\ , while
the scattered electrons in the case of \wes\ produce unquenched
scintillation light. In our fit to DAMA data for \wns\ we do include
channeling following ref.~\cite{Bernabei:2007hw} (similar as
in~\cite{Fairbairn:2008gz}), but we comment also on the case when no
channeling occurs.

\subsection{WIMP--electron scattering}
\label{sec:WEI}

To obtain an expression for the event rate in the case of \wes\ it is
necessary to take into account the fact that electrons are bound to the
atoms. The kinematics of scattering off bound electrons has some important
differences compared to scattering off free particles. The bound electron
does not obey the free-particle dispersion relation $E_{e\mathrm{(free)}}^2
= p^2 + m^2$. Instead it has a fixed energy $E_e = m_e - E_B$, determined by
the binding energy of the atomic shell, $E_B \ge 0$, whereas its momentum
$p$ follows a distribution which is given by the square of the Fourier
transform of the bound state wave function corresponding to that shell.
Energy conservation reads in this case $E_\chi + m_e - E_B = E_\chi' +
E_e'$, or
\be\label{eq:Eep}
E_e' = m_e + E_d - E_B \,.
\ee
After some tedious algebra one arrives at the following expression for $E_d$:
\be\label{eq:WE-Ed}
E_d \approx -\frac{p^2}{2 m_\chi} - p v \cos\theta \,,
\ee
where\footnote{We always denote the DM momentum with $\vect{k}$ and the
electron (or nucleus) momentum with $\vect{p}$, see
fig.~\ref{fig:4f-diagram}. Bold symbols refer to 3-vectors and $k \equiv
|\vect{k}|$, and similar for $p$.} $\cos\theta = \vect{k}\vect{p}/kp$ and we
used the approximation $E_d \ll m_e \le E_e \ll m_\chi$ and $v \sim
10^{-3}$. We see that to obtain detectable energies relevant for DAMA ($E_d$
of few keV), electron momenta of order MeV are required.

In appendix~\ref{app:WE} we give the details on the calculation of the scattering
cross section and count rate in the case of \wes, taking into account the
peculiarities of scattering on bound electrons. Here we only report the final results. 
Assuming the axial vector Dirac structure, $\Gamma_\chi = \Gamma_e = A$, as motivated above, the count rate is (we also set
$c^\chi_A = c^e_A = 1$ for simplicity):
\begin{align}
  \frac{dR^\wes}{dE_d} \simeq
  \frac{3 \rho_0 m_e G^2}{4\pi (m_{\rm I} + m_{\rm Na}) m_\chi} \,
    \sum_{nl} \sqrt{2 m_e (E_d - E_{B,nl})} \, (2l+1)
    \int\!\frac{dp\,p}{(2\pi)^3} \, |\chi_{nl}(p)|^2 \,
    I(v_{\rm min}^\wes) \,.
  \label{eq:dRdE-WEI}
\end{align}
Here $\chi_{nl}(p)$ is the momentum wave function of the electron, and the function $I(v_{\rm min})$ is
\begin{align}\label{eq:defI}
  I(v_{\rm min}) \equiv \int\!d^3v
    \frac{f_\odot(\vect{v})}{v} \, \theta(v - v_\mathrm{min}) \,,
\end{align}
while the minimal velocity required to give detectable
energy $E_d$ follows from eq.~\ref{eq:WE-Ed}:
\begin{align}\label{eq:vmin-wes}
  v_{\rm min}^\wes \approx \frac{E_d}{p} + \frac{p}{2 m_\chi} \,.
\end{align}
For $m_\chi \gtrsim 10$~GeV and $p$ of order MeV the first term dominates. 

The sum in eq.~\ref{eq:dRdE-WEI} is over the atomic shells of both iodine and sodium with quantum
numbers $nl$, and $E_{B,nl}$ is the corresponding binding energy. 
The
electron can only be kicked out of its atomic shell if its binding
energy is smaller than the total energy deposited in the detector (cf. eq. \ref{eq:Eep}):
\be\label{eq:EdEB}
E_d \ge E_{B,nl} \,.
\ee
Only the shells satisfying this requirement can contribute to the
event rate in eq.~\ref{eq:dRdE-WEI}. The momentum wave function
$\chi_{nl}(p)$ is defined in eq.~\ref{eq:psi-p} in appendix~\ref{app:WE}. 
Technical details on how we implement the wave function numerically are
given in appendix~\ref{app:wave-function}; the results for the iodine and
sodium wave functions are shown in fig.~\ref{fig:wf}. We see that the
dominant contribution to \wes\ scattering in DAMA comes from the inner
$s$-shells of iodine because these are largest at high $p$. Electrons from
the $1s, 2s, 2p$ shells 
--- depicted as thin light curves in fig.~\ref{fig:wf} --- do
not contribute to the DAMA signal region of $E_d \simeq 2-4$~keV 
since the binding energies are too large, respectively
$33.2$ keV,
$5.2$ keV, and $4.7$~keV \cite{Bearden:1967a}. 
The shell dominating the signal in the
2--4~keV region is the $3s$ shell of iodine, with a binding energy of about
1~keV. Apparently this has been overlooked in ref.~\cite{Bernabei:2007gr},
 while it has important consequences on the size of the needed cross section,
see discussion in sec.~\ref{sec:results-A}.

\subsection{WIMP--atom scattering}
\label{sec:WA}

Let us consider now the case when the electron on which the DM particle
scatters remains bound and the recoil is taken up by the whole atom.
According to the coordinate space Feynman rules, the matrix element for
\wa\ for an electron in atomic shell $nlm$ in the initial
state and $n'l'm'$ in the final state is given by
\begin{align}
  \mathcal{M}^{rr'ss'}_{nlm,n'l'm'} =  G \,
    \int\!d^3x \, \psi_{n'l'm'}^*(\vect{x}) \psi_{nlm}(\vect{x})  \, e^{-i \vect{p'} \vect{x}} \,
    e^{i (\vect{k} - \vect{k'}) \vect{x}} \,
    \bar{u}_\chi^{r'} \Gamma_\chi^\mu u_{\chi}^r \, \bar{u}_e^{s'} \Gamma_{e\mu} u_e^s \,.
  \label{eq:ME-WAE-1}
\end{align}
Here, $\vect{k}$ and $\vect{k'}$ are the initial and final momenta of the
WIMP, and $\vect{p'}$ is the average momentum of the electron in the final
state resulting from the motion of the whole atom after the scattering.
Since most of the recoil momentum is carried by the nucleus, $|\vect{p'}|$
is smaller than $|\vect{k} - \vect{k'}|$ by a factor of $m_e/m_N$, and can
therefore be neglected. The coordinate space wave function of the electron
in the state with orbital quantum numbers $nlm$ is denoted by
$\psi_{nlm}(\vect{x})$.  Again we specialize to the case of axial vector
coupling, $\Gamma_\chi^\mu = \Gamma_e^\mu = \gamma^\mu \gamma^5$ and set
$c^\chi_A = c^e_A = 1$. 
We use non-relativistic spinors, which is
certainly justified for $u_\chi^r$ and $u_\chi^{r'}$, and also for $u_e^s$
except, perhaps, for electrons from the $1s$ shell of iodine. In this last
case, relativistic corrections are of order 20\%. 

Let us first consider the case when the electron remains in its state, and
hence the scattering on the atom is elastic (\wae). Then we have $s=s'$ and
$nlm = n'l'm'$. Furthermore, we have to sum coherently over all shells and
electron spins, since it is impossible in principle to identify on which
electron the WIMP has scattered. It turns out that for the axial vector case
the spin sum $\sum_s \bar{u}_e^s \gamma^\mu\gamma^5 u_e^s$ vanishes. This
can be verified by using explicit expressions for the spinors $u_e^s$, and
follows from the fact that the different sign due to $\gamma_5$ of
right-handed and left-handed components of the electron cancel each other in
case of a coherent sum over spins.\footnote{This argument will not hold if
an unpaired valence electron is available so that we cannot sum over spins. However,
most chemically bound systems are formed in such a way that this does not
happen. 
Even in this case \wae\ would be suppressed since scattering on outer electrons is highly
suppressed by the smallness of the binding energy of these electrons compared to
the transferred momentum. 
}
The elastic scattering may be relevant for
other Lorentz structures where this cancellation does not occur. However, in
sec.~\ref{sec:ref} we have argued that the only case of practical relevance is
the axial coupling, and therefore we will not consider \wae\ further.

We are left now with the case where the electron is excited to an outer free
shell which corresponds to inelastic WIMP--atom scattering (\wai). In this
case the sum over all occupied electron states $nlm$, over all unoccupied
states $n'l'm'$, and over WIMP and electron spins has to be incoherent
because one can distinguish in principle different initial and final states,
e.g.\ by x-ray spectroscopy. The differential cross section in this case is
obtained as
\be\label{eq:sig-Ed}
  \frac{d \sigma^\wa}{d E_d} = 
  \frac{m_N\, \overline{|\mathcal{M}|^2}}{32\pi m_e^2 m_\chi^2 v^2}  \,.
\ee
Plugging in the matrix element from eq.~\ref{eq:ME-WAE-1} we get
\begin{align}
  \frac{d\sigma^\wai}{dE_d} &= \frac{3 m_N G^2}{2 \pi v^2} \,
    \sum_{nlm} \sum_{n'l'm'}   
    |\bra{n'l'm'} e^{i (\vect{k} - \vect{k'}) \vect{x}} \ket{nlm}|^2
  \label{eq:dsigma-WAI}
\end{align}
with 
\begin{align}
  \bra{n'l'm'} e^{i (\vect{k} - \vect{k'}) \vect{x}} \ket{nlm} &\equiv
    \int\!d^3x \, \psi_{n'l'm'}^*(\vect{x}) \psi_{nlm}(\vect{x}) \,
           e^{i (\vect{k} - \vect{k'}) \vect{x}} \,.
  \label{eq:ME-WAI-1}
\end{align}
The expression for the counting rate is obtained from eq.~\ref{eq:dRdE-1},
\begin{align}
  \frac{dR^\wai_N}{dE_d} = 
   \frac{m_N}{m_{\rm I} + m_{\rm Na}}
    \frac{3 \rho_0 G^2}{2\pi \, m_\chi} \,
    \sum_{nlm} \sum_{n'l'm'} 
    |\bra{n'l'm'} e^{i (\vect{k} - \vect{k'}) \vect{x}} \ket{nlm}|^2 \,
    I(v_{\rm min}^\wai) \,,
  \label{eq:dRdE-WAI}
\end{align}
with $N = \rm I, Na$. The function $I$ is defined in eq.~\ref{eq:defI}, and
the minimal velocity required to give detectable energy $E_d$ follows from
the kinematics implied by energy conservation, $E_d = E_\chi - E'_\chi =
\delta E_B + m_N v_N^2/2$, and momentum conservation, $\vect{k} = \vect{k}'
+ m_N \vect{v}_N$:
\begin{align}
  v_{\rm min}^\wai = \frac{E_d (m_\chi + m_N) - m_N \delta E_B}
                     {m_\chi \sqrt{2 m_N (E_d - \delta E_B)}} \,,
  \label{eq:vmin-WAI}
\end{align}
where $\delta E_B$ is the difference of the binding energies of the initial
and final shells: $\delta E_B = E_{B,nlm}-E_{B,n'l'm'}$. Details on how we
calculate the matrix elements involving the wave function in
eq.~\ref{eq:dRdE-WAI} are given in appendix~\ref{app:wave-function}.

\subsection{Loop induced WIMP--nucleus scattering}
\label{sec:loop-rate}

The event rate for loop induced DM--nucleus scattering is easy to obtain
from the differential cross sections ${d\sigma_N}/{dE_d}$ in
eqs.~\ref{eq:vector:nucl}--\ref{scalarDM:two-loop:diff} and the general
expression for the counting rate eq.~\ref{eq:dRdE-1}:
\be
\frac{dR_N^\wns}{dE_d}=\frac{\rho_0}{m_\chi} \, \frac{1}{m_I+m_{Na}}
\left(\frac{d\sigma_N}{dE_d} v^2\right) I(v_{\rm min}^\wns) \,.
\label{eq:dRdE-WNE-1}
\ee
The function $I$ is defined in eq.~\ref{eq:defI}, while the minimal velocity
to produce a detectable energy $E_d$ is given for WIMP--nucleus elastic
scattering by $v_{\rm min}^\wns = \sqrt{E_d m_N / 2 \mu_N^2}$ with $\mu_N =
m_\chi m_N / (m_\chi + m_N)$.

We now specialize to the $V\otimes V$ case. The event rate depends on the
$\chi$ mass and the coupling constant of the effective operator $G$ (we set
$c_V^\chi = c_V^\ell=1$ from now on). For easier comparison with previous
works, it is useful to trade $G$ for the total $\chi e\to \chi e$ cross
section $\sigma_{\chi e}^0 = {G^2 m_e^2}/{\pi}$, eq. \ref{eq:sig0ferm}. For
the $V\otimes V$ case we thus have
\be\label{eq:VVLL}
 \frac{d\sigma_N}{dE_d} v^2=\sigma_{\chi e}^0\times \frac{m_N}{18 m_e^2} 
 \Big(\frac{\alpha Z}{\pi}\Big)^2 F(q)^2
 \Big[\log\Big(\frac{m_\ell^2}{\mu^2}\Big)\Big]^2\,,
\ee
to be inserted in eq.~\ref{eq:dRdE-WNE-1}. As discussed in
appendix~\ref{app:loop} this leading log approximation is quite
accurate.  Nevertheless, in the numerical calculations we use the full
expressions given in appendix~\ref{app:loop}. We set $\mu = 10$~GeV,
since this corresponds roughly to the scale $\Lambda$, where our
effective theory is defined. Furthermore, we assume (somewhat
arbitrarily) equal couplings to all three leptons. The logarithm in
eq.~\ref{eq:VVLL} implies then a relative contribution of $e:\mu:\tau
\simeq 30:7:1$. 
Note that the rate is dominated by the contribution
from the electron in the loop assuming equal couplings at the scale
$\Lambda\sim 10\text{~GeV}$. 
Therefore, our results are
conservative, in the sense that per assumption DM has to couple to the
electron.

\section{Fit results for DAMA, CDMS, and XENON10}
\label{sec:results}

\subsection{Vector like interactions and loop induced WIMP--nucleon scattering}
\label{sec:results-V}

The event rate in the case of the loop induced \wns,
eq.~\ref{eq:dRdE-WNE-1}, is very similar to the corresponding expression for
usual spin-independent elastic WIMP scattering on the nucleus (see, e.g.,
refs.~\cite{Jungman:1995df, Fairbairn:2008gz, Savage:2008er}), denoted as
``standard case'' in the following. The main difference is the replacement
of the atomic mass number $A$ by the charge number $Z$ (and an additional
logarithmic $E_d$ dependence beyond the leading log approximation).
Therefore we expect that the fit of DAMA and the compatibility to CDMS and
XENON will be very similar to the standard case, see e.g.,
refs.~\cite{Chang:2008xa, Fairbairn:2008gz, Savage:2008er}.

\begin{figure}[t]
\centering 
\includegraphics[width=0.6\textwidth]{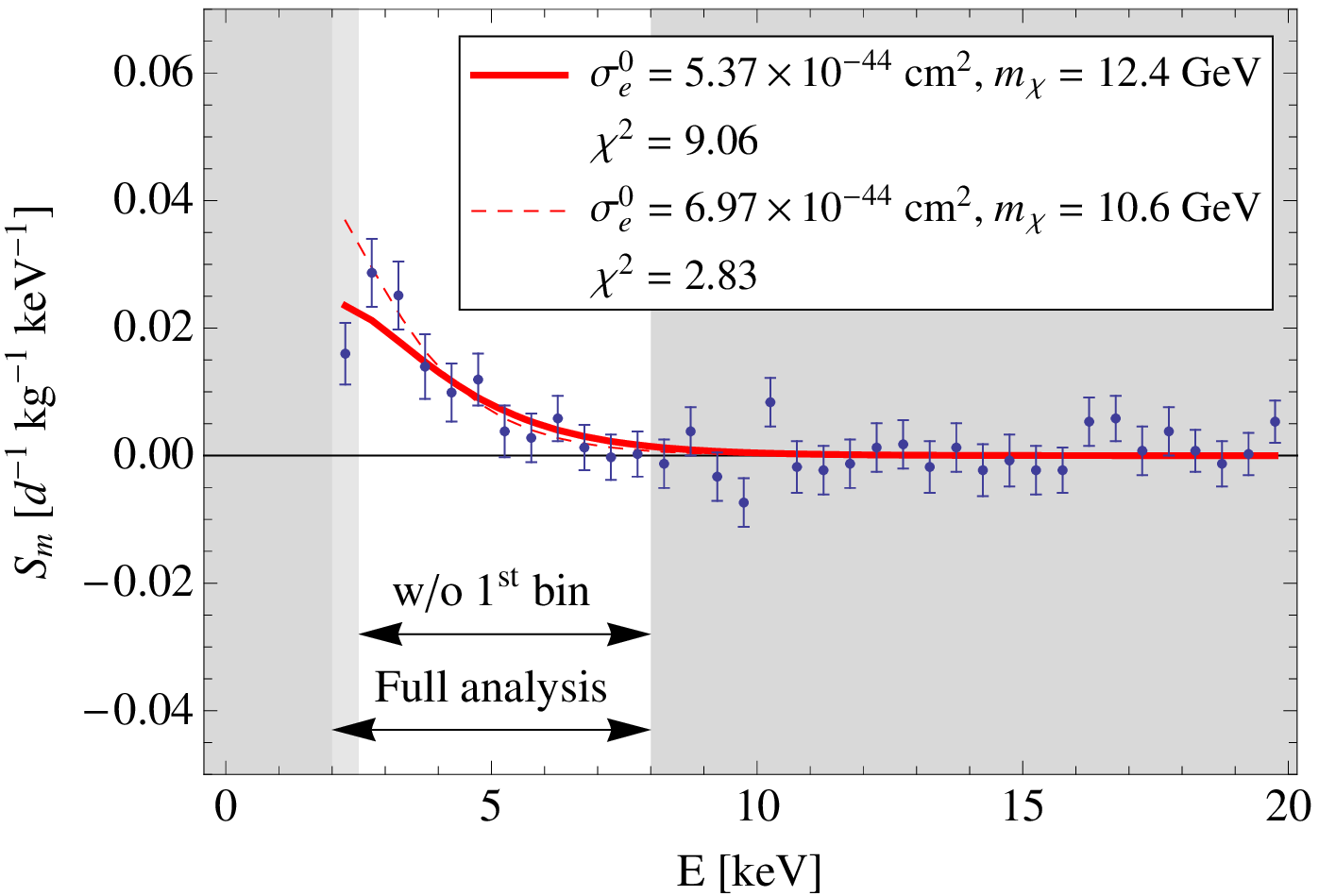}\\
\hspace*{4mm}\includegraphics[width=0.575\textwidth]{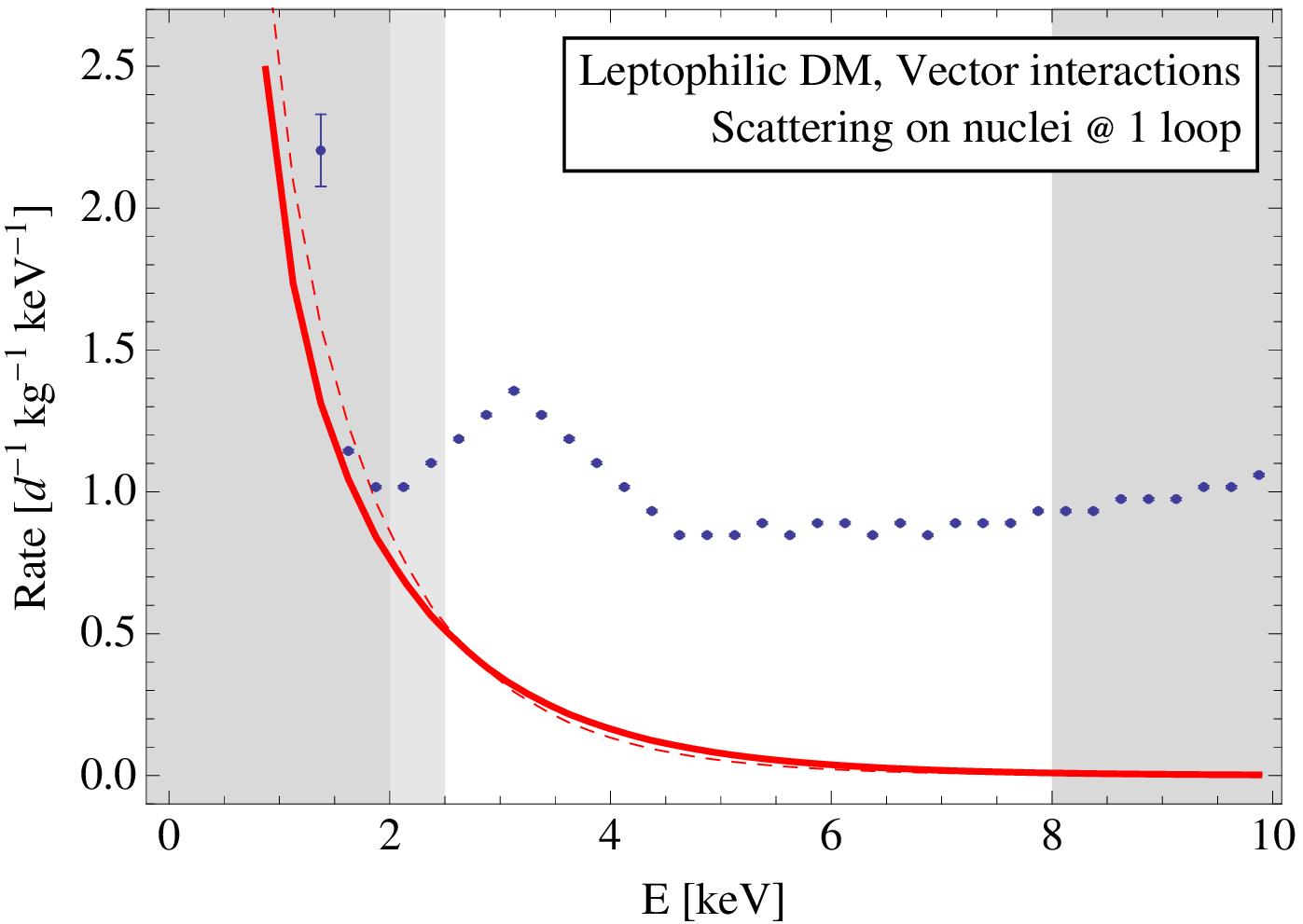} 
  \mycaption{\label{fig:best_fit-V} Predicted spectrum for the modulated
  (top) and unmodulated (bottom) event rate in DAMA at the best fit point
  assuming loop induced WIMP--nucleus scattering resulting from vector-like
  DM--lepton couplings. Results are shown for using all data points from
  2--8~keV (solid) and for omitting the 1st bin (dashed). The parameter
  values and the $\chi^2$-values are given in the legend.}
\end{figure}

Our numerical analysis of DAMA, CDMS, and XENON data follows closely
ref.~\cite{Fairbairn:2008gz} where technical details on the fit can be
found. For the DAMA fit we use the spectral data on the annual modulation
amplitude $S_m$ from the threshold of 2~keV up to 8~keV. The data above this
energy are consistent with no modulation and since our model does not
predict any features in that region they do not provide an additional
constraint on the fit, apart from diluting the overall goodness of fit. In
addition to the data on the amplitude of the modulated count rate we use
also the unmodulated event rate as a constraint in the fit. While the bulk
of these events will come from various unidentified backgrounds, every model
has to fulfill the constraint of not predicting more unmodulated events than
actually observed in DAMA.  Fig.~\ref{fig:best_fit-V} shows the predicted
spectrum at the best fit point compared to the DAMA data. Note that the
error bars on the unmodulated rate (lower panel) are hardly visible for most
of the data points, as a result of the huge number of events in DAMA.  For the
analysis using data from 2--8~keV the best fit point is at $m_\chi =
12.4$~GeV, $\sigma^0_{\chi e} = 4.5\times 10^{-44}\,{\rm cm}^2$, and we obtain
the excellent fit of $\chi^2/{\rm dof} = 9.1/10$. If we drop the first data
point the fit even improves to $\chi^2/{\rm dof} = 2.8/9$. 

\begin{figure}[t]
\centering \includegraphics[width=0.6\textwidth]{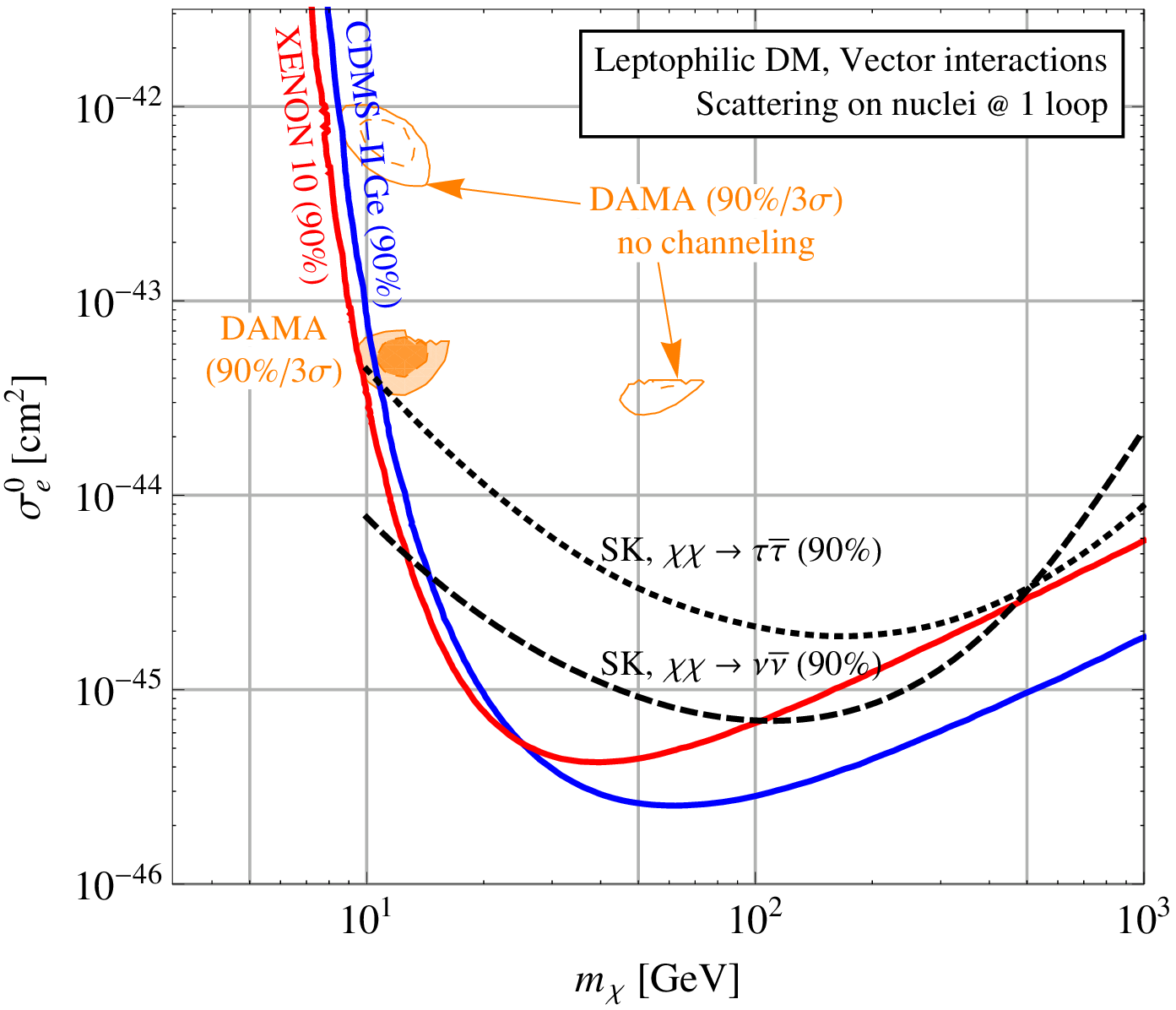} 
  \mycaption{\label{fig:msigma-V} DAMA allowed region at 90\% and
  3$\sigma$~CL in the case of 1-loop induced WIMP--nucleus scattering
  ($V\otimes V$ coupling) in the plane of the WIMP mass and the
  WIMP--electron cross section $\sigma^0_{\chi e} = G^2 m_e^2/\pi$. Regions
  are shown with and without taking into account the channeling effect.
  Furthermore, we show the bounds at 90\%~CL from CDMS-II and XENON10. The
  dashed curves show the 90\%~CL constraint from the Super-Kamiokande limit
  on neutrinos from the Sun, by assuming annihilation into $\tau\bar\tau$ or
  $\nu\bar\nu$, see sec.~\ref{sec:SK} for details.}
\end{figure}

The allowed regions in the plane of DM mass and scattering cross section are
shown in fig.~\ref{fig:msigma-V}.
For easier comparison with the case of
scattering off electrons we parameterize the cross section on the vertical
axis in terms of $\sigma_{\chi e}^0$, see eq.~\ref{eq:VVLL}. The spectral
data on the modulated signal results in an allowed region for rather small
DM masses around $m_\chi \simeq 12$~GeV. If channeling is not taken into
account an allowed region appears at similar DM masses but at higher cross
sections (due to scattering off sodium~\cite{Gondolo:2005hh}).\footnote{In
both cases (with and without channeling) there is also a local minimum
around a WIMP mass of about 80~GeV from unchanneled scatterings off iodine.
In fig.~\ref{fig:msigma-V} we show confidence regions defined with respect
to the global minimum, and this second region does not appear at 3$\sigma$
if channeling is included.}
This DAMA allowed region has to be compared to the constraints from CDMS and
XENON. The compatibility with these bounds is similar to the standard case:
while marginal compatibility might remain there is clearly severe tension
between the DAMA signal and the CDMS and XENON bounds in this framework.
Here we will not elaborate on this question further and refer to
refs.~\cite{Chang:2008xa, Fairbairn:2008gz, Savage:2008er} for detailed
discussions of the DAMA versus CDMS/XENON compatibility in the standard
case. 

The main motivation for considering electron interacting
DM---namely avoiding the constraints from nuclear recoil experiments,
is thus invalidated by the loop induced nucleon scattering. We now turn
to the axial-axial coupling, where the loop induced scattering can be
avoided.

\subsection{Axial vector like interactions and WIMP--electron scattering}
\label{sec:results-A}

\begin{figure}[t]
\centering 
\includegraphics[width=0.6\textwidth]{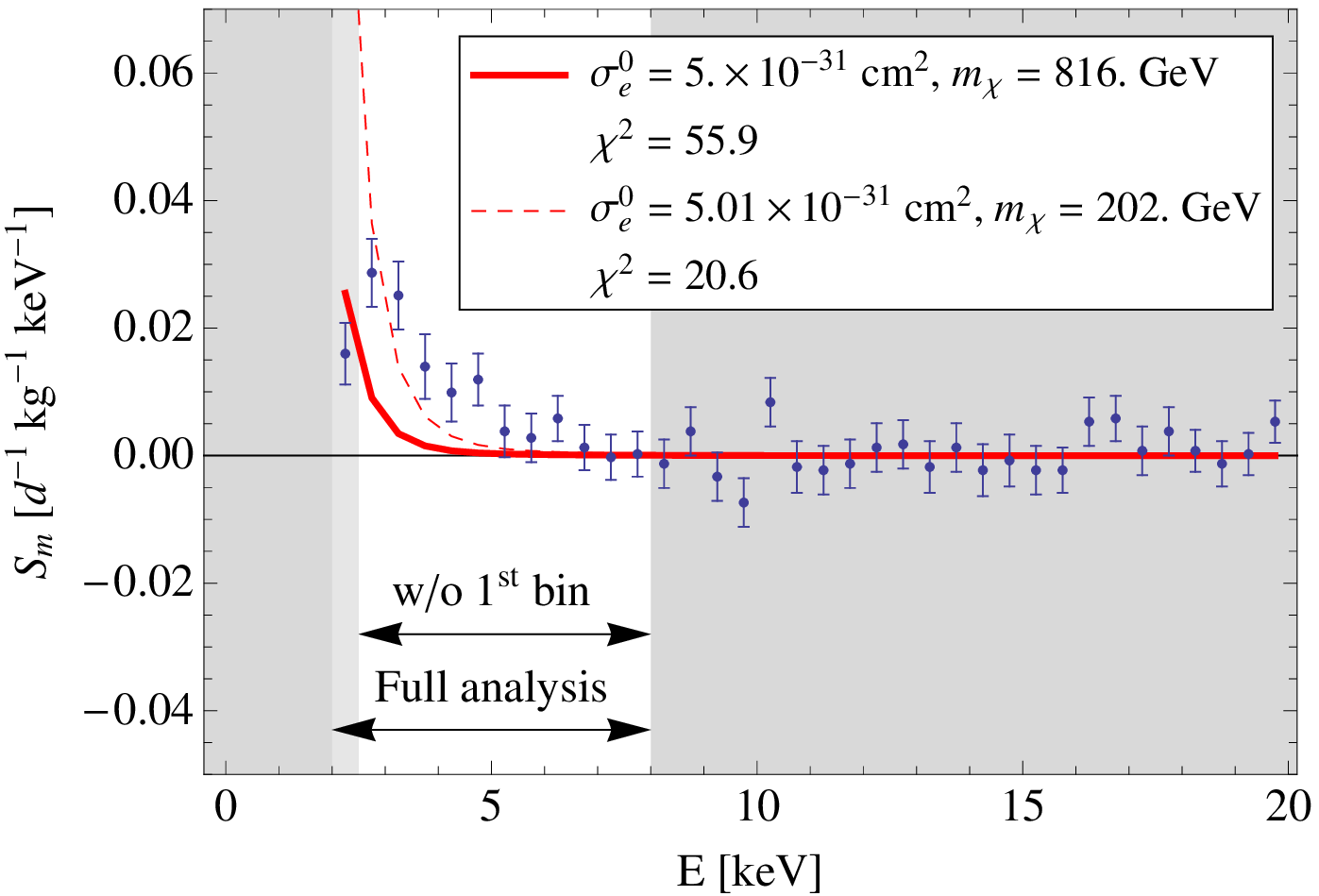}\\
\hspace*{4mm}\includegraphics[width=0.575\textwidth]{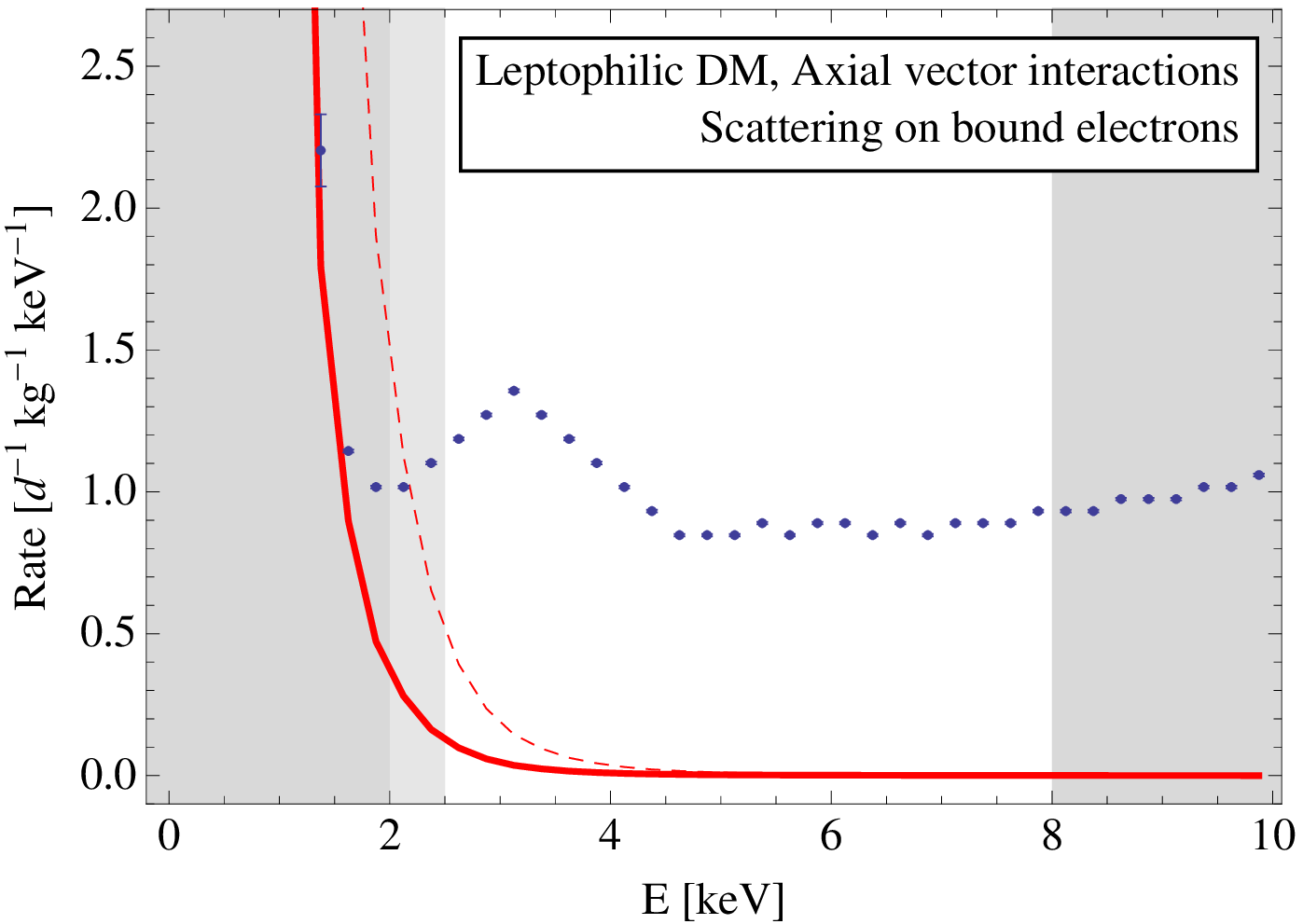}
  \mycaption{\label{fig:best_fit-A}Predicted spectrum for the modulated
  (top) and unmodulated (bottom) event rate in DAMA at the best fit point
  assuming WIMP--electron scattering resulting from axial vector-like
  DM--electron couplings. Results are shown for using all data points from
  2--8~keV (solid) and for omitting the 1st bin (dashed). The parameter
  values and the $\chi^2$-values are given in the legend.}
\end{figure}

We perform a similar fit to DAMA data as before but using now
eq.~\ref{eq:dRdE-WEI} for the event rate for \wes. The predicted modulated
and unmodulated DAMA event rates at the best fit in this case are shown in
fig.~\ref{fig:best_fit-A}. Using the data from 2--8~keV we obtain a rather
bad fit to the modulated spectrum with $\chi^2/{\rm dof} = 55.9/10$, which
corresponds to a probability of $2\times 10^{-8}$. The prediction drops too
fast with energy in order to provide a satisfactory fit to the
data. If we omit the first energy bin the fit improves considerably to
$\chi^2/{\rm dof} = 20.6/9$ corresponding to a probability of 1.4\%.  We
find, however, that the parameter values from this fit predict a very sharp
rise for the spectrum of the unmodulated event rate in DAMA, see lower panel
of fig.~\ref{fig:best_fit-A}. In the fit we have required that the
unmodulated prediction stays below the observed rate within the analysis
window down to 2~keV. However, DAMA shows also some data points for the
unmodulated rate below 2~keV, which are not compatible with the predicted
rate. While it is not possible to use data below 2~keV for the modulation,
it seems likely that they rule out models predicting more events than
observed. The \wes\ fit shown as dashed curve in fig.~\ref{fig:best_fit-A}
predicts more than a factor 3 more events than observed in the first two
bins below 2~keV, where error bars are still very small. We conclude that
\wes\ has severe problems to explain the spectral shapes of the modulated
and unmodulated components of DAMA data.

\begin{figure}[t]
\centering 
\includegraphics[width=0.6\textwidth]{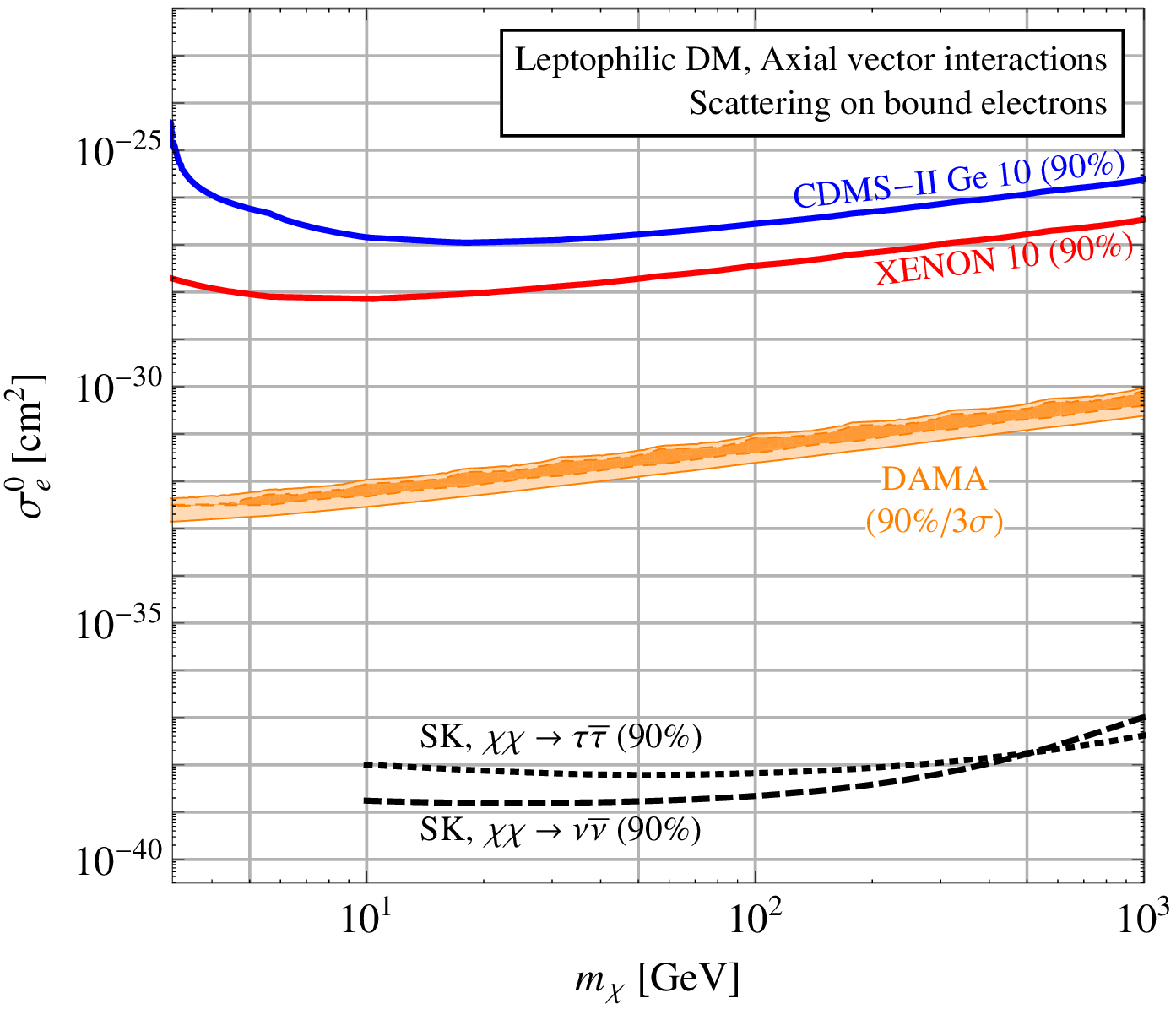}
  \mycaption{\label{fig:msigma-A} DAMA allowed region at 90\% and
  3$\sigma$~CL in the case of WIMP--electron scattering (axial-axial
  coupling) in the plane of the WIMP mass and the WIMP--electron cross
  section $\sigma^0_{\chi e} = G^2 m_e^2/\pi$. We stress that these regions
  are obtained with respect to the best fit point, which by itself does not
  provide a satisfactory fit to DAMA modulated and unmodulated spectral
  data, see fig.~\ref{fig:best_fit-A}. We show also the bounds at 90\%~CL
  from CDMS-II and XENON10 from inelastic WIMP--atom scattering. The dashed
  curves show the 90\%~CL constraint from the Super-Kamiokande limit on
  neutrinos from the Sun, by assuming annihilation into $\tau\bar\tau$ or
  $\nu\bar\nu$, see sec.~\ref{sec:SK} for details.}
\end{figure}

If we ignore the problems of the spectral fit and despite the low
goodness-of-fit consider ``allowed regions'' in the plane of WIMP mass and
cross section relative to the best fit point we obtain the results shown in
fig.~\ref{fig:msigma-A}. We observe that very large cross sections are
required:
\be\label{eq:Xs-wes}
\sigma^0_{\chi e} \sim 10^{-31}\,\rm cm^2 \times \bfrac{m_\chi}
{100 \,\rm GeV} \,,
\ee
where the linear dependence on $m_\chi$ holds for $m_\chi \gtrsim 10$~GeV.
The vastly different best fit cross sections for \wns\ and \wes\ follow from
the discussion in sec.~\ref{sec:signals} where we estimated the relative
size of the corresponding counting rates, see eq.~\ref{eq:hierarchy}. Here
we do not explore other phenomenological consequences of such a large cross
section. Just note from eq.~\ref{eq:sig0ferm} that we can realize a cross
section of this order of magnitude only with a relatively low scale for the
new physics of $\Lambda \lesssim 0.1$~GeV, where we have assumed for the
couplings $c_A^\chi\sim c_A^e \sim 1$. If in a particular model constraints
on the coupling constant $c_A^e$ apply (see, for
example~\cite{Fayet:2007ua}), $\Lambda$ has to be accordingly smaller. Note
that the momentum transfer for \wes\ is given by the momentum of the bound
electron, which has to be of order MeV. This provides a lower bound on
$\Lambda$ in order to describe the interaction by using the effective
theory.

The cross section from eq.~\ref{eq:Xs-wes} is about 5 orders of magnitude
larger than the result obtained in ref.~\cite{Bernabei:2007gr}, which finds
pb-size $\sigma^0_{\chi e}$ at $m_\chi \sim 100$~GeV. Let us comment,
therefore, on the differences of our analysis to the one
from~\cite{Bernabei:2007gr}. Apparently the main difference is that we take
into account the special kinematics related to the scattering off bound
electrons, whereas in ref.~\cite{Bernabei:2007gr} electrons are treated as
effectively free with a momentum distribution obtained from the wave function.
Our calculation outlined in sec.~\ref{sec:WEI} and appendix~\ref{app:WE} leads
to several suppression factors in the \wes\ event rate with respect to the
expression used in~\cite{Bernabei:2007gr}. Our minimal velocity $v_{\rm
min}^\wes$ from eq.~\ref{eq:vmin-wes} is roughly a factor two larger than the
one used in~\cite{Bernabei:2007gr} requiring to go further out in the tail of
the WIMP velocity distribution.  Furthermore, the condition eq.~\ref{eq:EdEB},
$E_d \ge E_B$, which prevents the contribution of the inner shells of iodine,
has not been imposed in~\cite{Bernabei:2007gr}. From fig.~\ref{fig:wf} we see
that at $p\sim 1$~MeV the wave function of the iodine $1s$ shell is about 2
orders of magnitude larger than the one of the $3s$ shell, which actually gives
the first relevant contribution after requiring that the binding energy has to
be lower than $E_d \sim$~few keV. 

\bigskip

As mentioned above, in the case under consideration inelastic \wa\ may
contribute to experiments searching for nuclear recoils. To calculate this
effect we would have to perform the sum in eq.~\ref{eq:dRdE-WAI} over all
occupied shells $nlm$ and all free shells $n'l'm'$. It turns out numerically
that transitions from $s$-shells to $s$-shells give the largest contributions,
see also appendix~\ref{app:wave-function}. In order to estimate the order of
magnitude we have taken into account transitions from the $1s, 2s, 3s$
orbitals to the first two free $s$-orbitals of the germanium and xenon
nuclei relevant in CDMS and XENON, respectively. In fig.~\ref{fig:msigma-A}
we show the constraints resulting from this estimate of the \wai\ event
rate. Numerically these constraints turn out to be very weak and the limits
are several orders of magnitude above the region indicated by DAMA; the good
sensitivities of CDMS and XENON to nuclear recoils cannot compensate the
large suppression of the \wai\ count rate compared to \wes, as estimated in
eq.~\ref{eq:hierarchy}.

Although the poor quality of the fit in the case of \wes\ already disfavors
this mechanisms as an explanation for the DAMA modulation signal, we will
show in the next section that constraints on neutrinos from DM annihilations
inside the Sun are even more stringent and exclude the cross sections
required for DAMA by many orders of magnitude if DM annihilations provide
neutrinos in the final states.

\section{Neutrinos from DM annihilations inside the Sun}
\label{sec:SK}

Any DM candidate has to fulfill the constraints on the upward through-going
muons coming from water Cerenkov detectors, like
Super-Kamiokande~\cite{Desai:2004pq}, and from neutrino
telescopes~\cite{Ackermann:2005fr, Achterberg:2006md, Abbasi:2009uz}.  Some
recent papers~\cite{Hooper:2008cf, Feng:2008qn} have discussed the
constraints on the DAMA region in the framework of standard WIMP--nucleus
scattering. Here, we reanalyze the bound coming from the Super-Kamiokande
experiment in the framework of leptonically interacting DM.

One important ingredient for the prediction of the neutrino flux coming from
DM annihilations inside a celestial body is the capture rate $C_\odot$,
which is proportional to the DM scattering cross section, see,
e.g.~\cite{Jungman:1995df}. In the calculation of this quantity, usually,
the WIMPs are assumed to interact with material at zero temperature,
neglecting the solar temperature of about $1.5\times 10^7$~K in the center
and $8.1\times 10^4$~K at the surface.  Although this is a reasonable
assumption for WIMP candidates interacting with hydrogen and the other
nuclei inside the Sun, it fails for the case of DM scattering on the free
electrons in the Sun. The effect of non-zero temperature on the capture rate
depends on the ratio of the thermal velocity of the target to the WIMP
velocity. The thermal kinetic energy $k_B T$ is independent of the mass, but
the thermal velocity is larger by a factor $\sqrt{m_p/m_e} \simeq 45$ for
electrons compared to hydrogen. 

\begin{figure}[t]
\centering 
\includegraphics[width=0.6\textwidth]{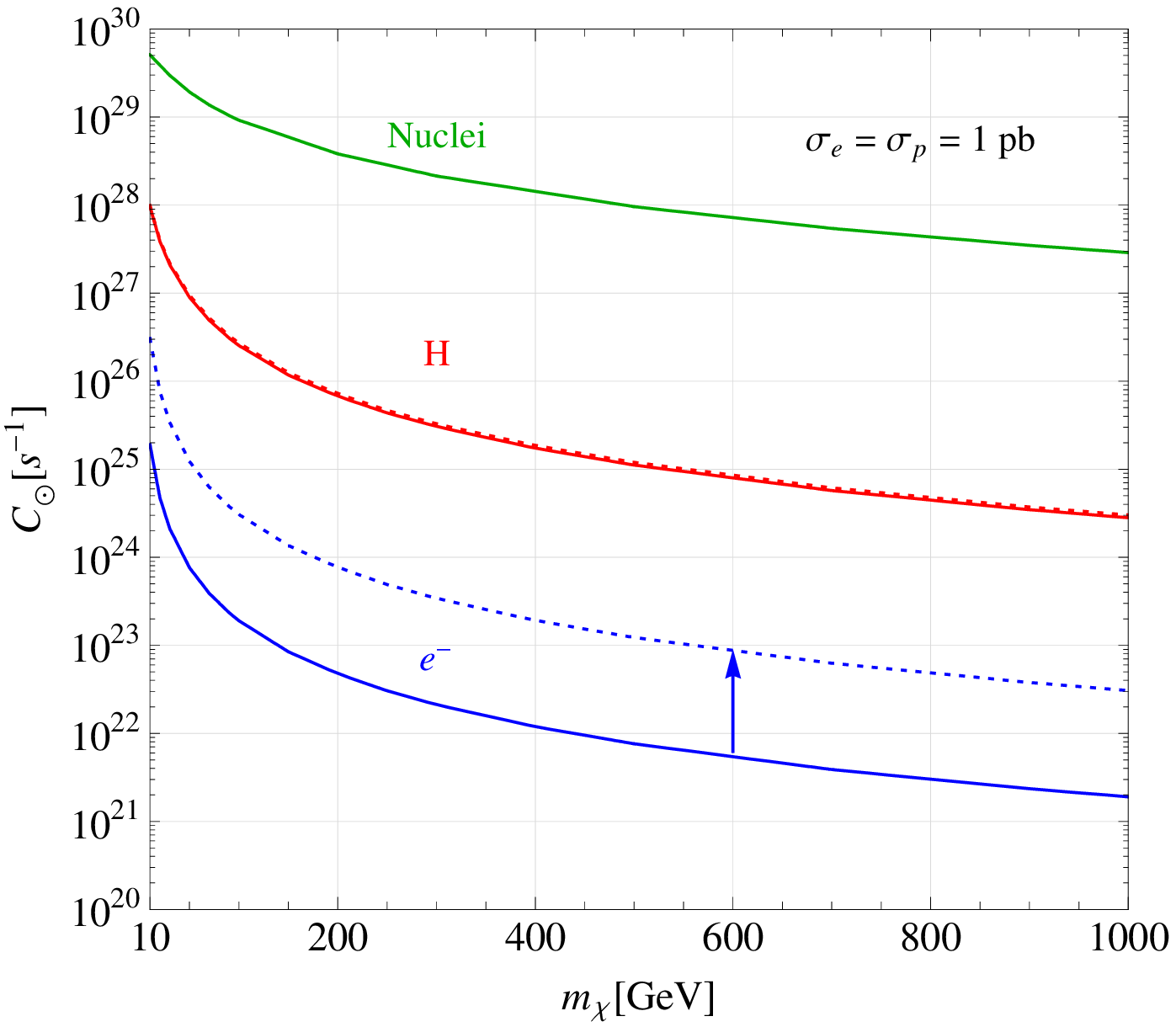}
  \mycaption{\label{fig:temp} WIMP capture rate in the Sun as a function of
  the WIMP mass assuming DM scattering off electrons, hydrogen, and all
  other nuclei in the Sun, with a scattering cross section of $10^{-36}\,\rm
  cm^2$. The solid curves correspond to scattering off particles at zero
  temperature, whereas the dotted curves show the effect of the actual
  temperature distribution inside the Sun~\cite{Bahcall:2004pz} for
  electrons and hydrogen.}
\end{figure}

We calculate the rate for WIMP capture by a body at finite temperature
following ref.~\cite{Gould:1987ir}, where the expression given there has to
be extended to include the motion of the Sun with respect to the DM halo.
The temperature distribution for the electrons inside the Sun is taken from
the solar model BS2005-AGS,OP~\cite{Bahcall:2004pz}. Fig.~\ref{fig:temp}
shows the effect of the non-zero temperature on the capture rate for
electrons, hydrogen and all other nuclei in the Sun. We find that the
capture rate on electrons is enhanced by about one order of magnitude,
while the effect is hardly visible at the scale of the plot for hydrogen.
The temperature effect can be neglected for scattering off heavier nuclei,
which dominates the capture in the case of loop induced \wns. In this case
one has to include also a suppression due to the nuclear form factor.  
Furthermore, we neglect the effect of WIMP evaporation, important only for
DM masses lower than 10~GeV~\cite{1987ApJ...321..560G} and the gravitational
effects from planets like Jupiter~\cite{Peter:2009mk}.

The annihilation rate $\Gamma_{\odot}$ is related to the capture rate
$C_{\odot}$ by~\cite{Jungman:1995df}
\be\label{eq:equil}
\Gamma_{\odot}=\frac{C_{\odot}}{2}\,\tanh^{2}\left(\frac{t_{\odot}}{\tau}\right)\,,\qquad
\tau = \frac{1}{\sqrt{C_{\odot}\,C_{A}}}\,,
\ee
where $t_{\odot} \simeq $ 4.5 Gyr is the age of the Sun and the parameter
$C_{A}$ depends on the WIMP annihilation cross section and on the effective
volume of the confining region in which the DM particles are trapped:
$C_{A}=\langle\sigma_{\rm{ann}}v \rangle\,V_{2}/V^{2}_{1}$ with
$V_{j}=6.6\times10^{28}\,(j\,m_{\chi}/\rm{10}\,\rm{GeV})^{-3/2}\,cm^{3}$. We
denote by $\langle\sigma_{\rm{ann}}v \rangle$ the thermally averaged total
annihilation cross section times the relative velocity, at the present time.
Capture and annihilation are in equilibrium if $\tau \ll t_\odot$. Then the
annihilation rate is just half the capture rate and becomes independent of
the annihilation cross section $\langle\sigma_{\rm{ann}}v \rangle$. For our
calculations we assume this limit, we comment on its validity in
appendix~\ref{app:equil}.
 
The neutrino flux at the detector from the annihilation channel $f$ with
branching ratio $BR_{f}$ is given by
\be
\frac{d \phi^{f}_{\nu}}{d E_{\nu}}=BR_{f}\frac{\Gamma_{\odot}}{4 \pi d^{2}}\frac{d N^{f}_{\nu}}{d E_{\nu}}\,,
\ee
with $d$ being the distance between the Earth and the Sun. Here we are
interested in annihilations into leptons. We consider the following four
channels: $\tau\bar{\tau}$, $\nu_{e}\bar{\nu}_{e}$,
$\nu_{\mu}\bar{\nu}_{\mu}$, and $\nu_{\tau}\bar{\nu}_{\tau}$. Note that
annihilations into electrons do not provide neutrinos, and muons are always
stopped before decay, giving rise to neutrinos in the MeV energy range which
is below the Super-Kamiokande threshold~\cite{Ritz:1987mh}. In the case of
direct neutrino channels, the initial neutrino spectrum is simply a Dirac
$\delta$ function centered at $E_{\nu}=m_{\chi}$, and we assume a
flavor-blind branching ratio, i.e., $BR_{\nu_{e}\bar{\nu}_{e}}=
BR_{\nu_{\mu}\bar{\nu}_{\mu}} = BR_{\nu_{\tau}\bar{\nu}_{\tau}} = 1/3$.
The results do not depend strongly on this assumption, since flavors are mixed
due to oscillations.\footnote{There is some difference of the
$\nu_\tau\bar\nu_\tau$-channel due to $\tau$ regeneration effects, which are
important for high energies. Assuming annihilations with branching ratios
equal to one for each of the three flavors we find that the muon neutrino
flux at the Earth is practically the same for all three initial flavors up
to $m_\chi \simeq 100$~GeV. For $m_\chi = 1$~TeV the ratio of the muon
neutrino fluxes at Earth is roughly $1:3.5:6.4$ for annihilations into
$\nu_e\bar\nu_e : \nu_\mu\bar\nu_\mu : \nu_\tau\bar\nu_\tau$. } For the
$\tau \bar{\tau}$ channel, we use the initial neutrino spectrum given
in~\cite{Cirelli:2005gh}. 

The neutrino spectrum $d N^{f}_{\nu}/d E_{\nu}$ at the detector is calculated
considering the effect of neutrino oscillation, coherent MSW matter effect,
absorption, and regeneration (see e.g., \cite{Cirelli:2005gh,
Blennow:2007tw}) by solving numerically the evolution equations of the neutrino
density matrix within the Sun. The neutrino oscillation parameters and mass
squared differences are fixed to the best fit values reported in
\cite{Schwetz:2008er}. We set $\theta_{13}$ to zero, avoiding in this way
possible Earth matter effects. 

The total muon flux is given by the formula, see e.g.~\cite{Barger:2007xf}:
\be
\Phi_{\mu} = \int^{m_{\chi}}_{E^{th}_{\mu}} d E_{\mu}
             \int^{m_{\chi}}_{E_{\mu}} d E_{\nu_{\mu}}
              \,\frac{d \phi^{f}_{\nu}}{d E_{\nu}}\,
              N_{A}\,R_{\mu}(E_{\mu},E^{th}_{\mu})\,
              \left[\mathcal{N}_{p}~\frac{d\sigma^{p}_{\nu}}{dE_{\mu}}(E_{\nu},E_{\mu})+
                    \mathcal{N}_{n}~\frac{d\sigma^{n}_{\nu}}{dE_{\mu}}(E_{\nu},E_{\mu})\right]\,,
  \label{eq:MuonFlux}
\ee
where $\mathcal{N}_{p}$ and $\mathcal{N}_{n}$ are the fractional number of
protons and neutrons at the point of muon production. We consider
$\mathcal{N}_{p} \simeq \mathcal{N}_{n} \simeq 0.5$, since for the
through-going muons in the Super-Kamiokande detector the interaction
can be assumed to occur in standard rock for which the number of protons is almost
equal to the number of neutrons ($Z=11, A=22$). $N_{A}$ is Avogadro's
number, and the effective number of nucleons per unit volume is given
by $N_{A}\,\rho_\mathrm{rock}$, with $\rho_\mathrm{rock}$ the density of the material.
The muon range $R_\mu$ is defined as the distance traveled by a muon
with initial energy $E_{\mu}$ and final energy equal to the
detector energy threshold $E^{th}_{\mu}$:
\be
  R_{\mu}(E_{\mu},E^{th}_{\mu}) 
         = \frac{1}{\beta} \,\ln \left(
    \frac{\alpha+\beta E_{\mu}}{\alpha+\beta E^{th}_{\mu}}\right)\,
  \label{eq:MuonRange}
\ee
with $\alpha \simeq 2.2 \times 10^{-3}\,
\rm{GeV}/\left({g}\,\rm{cm}^{-2}\right)$ and $\beta \simeq 4.4 \times
10^{-6}\,/\left(\rm{g}\,\rm{cm}^{-2}\right)$. The muon energy threshold has
been fixed to $E^{th}_{\mu}=1.6$ GeV, corresponding to the 7~m
path-length cut applied on upward through-going muons in the
Super-Kamiokande detector.
For the differential cross sections $d\sigma^{p,n}_{\nu}/dE_{\mu}$, we use
the analytic expressions for deep inelastic scattering given e.g.\ in
ref.~\cite{Barger:2007xf}.

Super-Kamiokande gives 90\%~CL limits on the muon flux induced by neutrinos
from DM annihilations in the Sun for cones of different opening angles
around the direction of the Sun~\cite{Desai:2004pq}. To be
conservative we use the limit for a cone with half-angle of $20^\circ$,
which should include 90\% of all muons at $m_\chi \simeq
18$~GeV~\cite{Desai:2004pq}, and a fraction approaching 100\% for larger
WIMP masses. The corresponding limit is $\Phi_{\mu} \le 1.1 \times
10^{-14}\,\rm{cm}^{-2}\,\rm{s}^{-1}$~\cite{Desai:2004pq}. Using this upper
bound on $\Phi_{\mu}$ we obtain via eq.~\ref{eq:MuonFlux} an upper bound on
the DM scattering cross section as a function of $m_\chi$.

These bounds are shown in figs.~\ref{fig:msigma-V} and \ref{fig:msigma-A} for
the case of loop induced WIMP--nucleus scattering (\wns) and WIMP--electron
scattering (\wes), respectively. We show the limit for annihilations into
$\tau\bar\tau$ and $\nu\bar\nu$ (assuming equal branchings into the 3 flavors).
In the case of \wns, annihilations into neutrinos exclude the region compatible
with DAMA, while annihilations into tau leptons might be marginally consistent
with it at $3\sigma$. In contrast, in the case of \wes\ the neutrino bound
excludes the region indicated by DAMA by more than 6 orders of magnitude. This
implies that if DM couples to electrons with a cross section as large as
indicated by the \wes\ DAMA fit, c.f.\ eq.~\ref{eq:Xs-wes}, DM annihilation
into neutrinos must be very strongly suppressed. This will be hard to achieve
because annihilation into charged leptons generates almost model independently
also annihilation into neutrinos from $W$ exchange at 1-loop. Thus annihilation
into neutrinos is typically suppressed by a loop factor of $O(10^{-4})$ compared
to annihilation into charged leptons. This rules out all
leptophilic DM models with dominant direct annihilation into leptons as an
explanation of DAMA/LIBRA.

\section{Conclusions}
\label{sec:concl}

In this study we have considered the hypothesis that DM has tree level
couplings only to leptons but not to quarks. Such a model has been proposed
in ref.~\cite{Bernabei:2007gr} to reconcile the DAMA annual modulation
signal with constraints from searches for nuclear recoils from DM
scattering. Our results imply, however, that this is not possible for the
following reasons:

\begin{enumerate}
\item
  By closing the lepton legs to a loop, we obtain a coupling to the charge
  of the nucleus by photon exchange. Whenever the Dirac structure of the
  DM--lepton coupling allows such a diagram at 1 or 2-loop WIMP--nucleus
  scattering will dominate over the scattering rate from the direct coupling
  to electrons, because the latter is highly suppressed by the high momentum
  tail of the bound state wave function. This leads to a situation very
  similar to the standard WIMP case, implying the well-known tension between
  DAMA and the constraints from CDMS and XENON10, see
  fig.~\ref{fig:msigma-V}.
\item 
  If the DM--lepton coupling is axial vector like, no loop will be induced
  and hence the scattering proceeds only by the interaction with electrons
  bound to the atoms of the detector. We have performed a careful analysis
  of this case, taking into account the peculiarities of scattering off
  electrons in bound states. We find that this model is strongly
  disfavored as an explanation of the DAMA signal because 
  \begin{enumerate}
  \item
    the predicted spectral shape of the modulated and/or unmodulated signal
    in DAMA provides a very bad fit to the data as shown in
    fig.~\ref{fig:best_fit-A}, and
  \item
    the cross section required to explain the DAMA signal is ruled out by
    Super-Kamiokande constraints on neutrinos from DM annihilations in the
    Sun, see fig.~\ref{fig:msigma-A}.
  \end{enumerate}
\end{enumerate}

The arguments 1 and 2(a) are rather model independent, relying only on the
presence of the effective DM--lepton vertex, while the argument in 2(b)
depends on the assumption that neutrinos are produced by DM annihilations.
Due to SU(2)$_L$ gauge symmetry, generically one expects that DM will couple
to both, charged leptons and neutrinos, which would open the annihilation
channel into $\nu\bar\nu$. If for some reason DM couples only to charged
leptons, DM would generically also annihilate into $\tau\bar\tau$, leading
again to the neutrino signal. In order to evade the Super-Kamiokande
constraint one has to forbid the coupling of DM to neutrinos and to the tau
lepton. Let us mention that the most generic way to avoid coupling to
neutrinos is the chiral coupling only to right-handed leptons. Note,
however, that such a chiral $V+A$ coupling involves a vector-like coupling
which will induce DM--quark scattering via the loop diagram and argument 1
applies.
Another way to evade the bound from annihilations would be to assume that DM
is not self-conjugate and postulate the presence of a large $\chi-\bar\chi$
asymmetry in our halo, see e.g., refs.~\cite{Nussinov:1985xr, Barr:1990ca,
Kaplan:1991ah}.

In conclusion, we have shown that the hypothesis of DM--interactions only with
leptons does not provide a satisfactory solution to reconcile the DAMA
annual modulation signal with constraints from other direct detection
experiments.

\bigskip
{\bf Note added:}
After the completion of this work we became aware of Ref.~\cite{Ahmed:2009rh},
where CDMS publishes constraints on electron-like events above 2~keV in
their detector. These results apply to the case of axial coupling, where
scattering off electron dominates. From Fig.~6 we find that our fit
predicts an unmodulated rate of 0.4 (1.5) events/d/kg/keV at 2~keV if the
lowest energy bin of the modulated rate is (is not) taken into account.
CDMS observes $1.93\pm 0.24$ events/d/kg/keV at 2~keV. Assuming a flat
background in the energy range of interest an upper limit on a possible
signal from DM of 0.5~events/d/kg/keV at 90\%~CL is obtained at 2~keV
(see Fig.~3 of \cite{Ahmed:2009rh}). The signal in Ge is expected to be roughly
one order of magnitude smaller than in iodine due to wave function
suppression. This estimate suggests that the results of \cite{Ahmed:2009rh} do
not rule out the DAMA region shown in Fig.~7. A more detailed analysis
may still be of interest, though.

\section*{Acknowledgments}
We would like to thank Alexander Merle, Thomas E.\ J.\ Underwood, and Andreas Weiler for
discussions, and Nicolao Fornengo and Marco Cirelli for help in the
calculation of the neutrino spectra.  This work was partly supported by the
Sonderforschungsbereich TR~27 of the Deutsche Forschungsgemeinschaft.  JK
received support from the Studienstiftung des Deutschen Volkes.

\appendix

\section{Loop induced DM--quark interactions}
\label{app:loop}

In this appendix we calculate the cross sections for DM--nucleon scattering
through the loop-induced interactions shown in fig.~\ref{fig:loop-diagram}.  The
main results were already collected in subsection~\ref{sec:loop} in the 
leading log approximation, while here we give 
full 1-loop results and describe how the approximate 2-loop results
were obtained. For calculations we use the {\tt FeynCalc} package \cite{Mertig:1990an}. The cross
section for scattering of a non-relativistic DM particle $\chi$ with mass
$m_\chi$ on a nucleus at rest carrying charge $Z$ and having a mass $m_N$ is
\be 
\frac{d\sigma}{dE_d}=\frac{\overline{|\mathcal{M}|^2}}{32 \pi m_N
m_\chi^2 v^2} \,,
\ee 
with $v\sim 10^{-3}$ the $\chi$ velocity, $E_d$ the kinetic recoil energy of
the nucleus and $\mathcal{M}$ the matrix element for $\chi N\to \chi N$
scattering.  

We start with the vector type interaction between leptons and DM, 
${\cal
L}_\ell = G (\bar \chi \Gamma_\chi^\mu \chi)(\bar \ell
c_V^\ell\gamma_\mu \ell)$ with
$\Gamma_\chi^\mu=(c_V^\chi+c_A^\chi\gamma_5)\gamma^\mu$, see
eq.~\ref{eq:4fermi}. The matrix element for $\chi N\to \chi N$ scattering
generated through the one loop diagram of fig.~\ref{fig:loop-diagram} is
then 
\be
\begin{split}\label{M:vector}
\mathcal{M} 
&=\,{\cal C}_V^{(1)}(\mu) \big(\bar u_\chi' \Gamma_\chi^\mu u_\chi \big) 
\langle N(p')| \sum_i Q_i \big(\bar q_i  \gamma_\mu q_i\big)|N(p)\rangle \\
& =\,{\cal C}_V^{(1)}(\mu)\big(\bar u_\chi' \Gamma_\chi^\mu u_\chi \big) 
Z F(q) \big(\bar u_N'  \gamma_\mu u_N\big) \,.
\end{split}
\ee
The sum is over the light quarks $q_i$ with charges $Q_i$, $F(q)$ is the
nuclear form factor defined in sec.~\ref{sec:loop}, and ${\cal
C}_V^{(1)}(\mu)$ is the 1-loop factor calculated in the $\overline{\rm MS}$ scheme
\be
{\cal C}_V^{(1)}(\mu)=
\frac{2 \alpha_{\rm em}}{\pi}G c_V^\ell \int_0^1 dx x(1-x) \log\Big[\frac{-x(1-x)q^2+m_\ell^2-i0}{\mu^2}\Big] \,,
\ee\label{CV1}
where $q^2\simeq -\kappa^2=-2 m_N E_d$ is the momentum transfer (in the calculation of the Super-Kamiokande bounds
we used $q^2 \simeq - \mathcal{O}(m_\chi^2 v^2)$). In the calculation we set $\mu=\Lambda$, with $\Lambda \sim 10$~GeV, because this
is the scale at which the Wilson coefficient $G$ is generated.\footnote{This
choice of $\mu$ does not minimize the size of the logarithm in ${\cal
C}_V^{(1)}(\mu)$. However, since the expansion parameter $\alpha_{\rm em}$
is small this does not invalidate the use of perturbation theory.  For a
choice of $\mu\ll \Lambda$ one would need to take into account
renormalization group flow and mixing of operators.} For $m_\ell \gg \kappa$ one can neglect the momentum transfer
in the above integral, giving an approximate expression
\be
{\cal C}_V^{\rm LL}(\mu)= \frac{\alpha_{\rm em}}{3\pi}G
c_V^\ell\log\big(m_\ell^2/\mu^2\big)\,, 
\ee
which is very precise for muon and tau running in the loop. It is  quite precise also for the electron, even though $m_e\sim \kappa$. The reason is that 
there is still a hierarchy $m_e \ll \mu\simeq\Lambda$. Neglecting the difference between $m_e$ and $\kappa$ then corresponds to a leading log (LL) approximation in the renormalization group running, while the induced error is only logarithmic in $1+\mathcal{O}(\kappa/m_e)$.

\begin{figure}
  \begin{center}
   \includegraphics[width=0.48\textwidth]{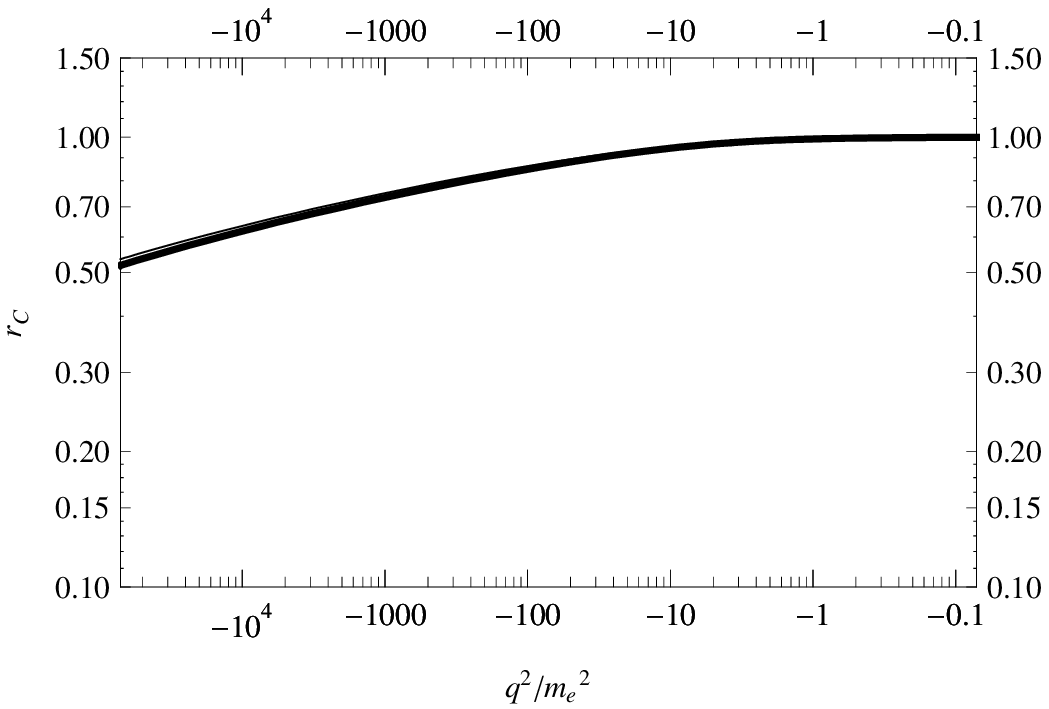}
    \includegraphics[width=0.48\textwidth]{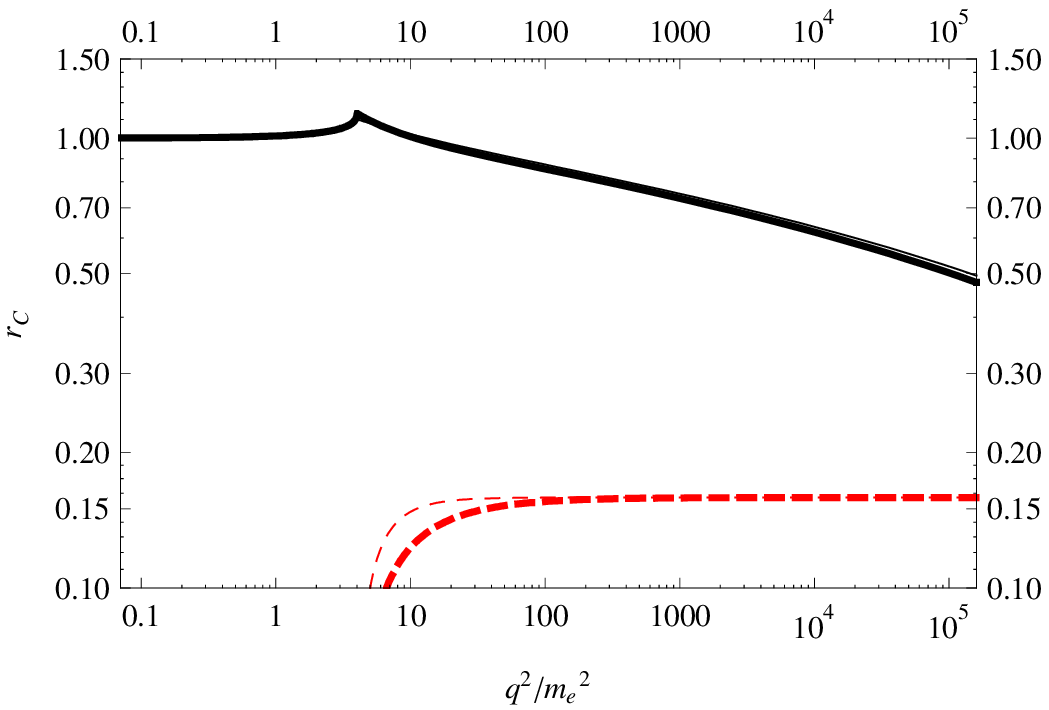}
  \end{center}
  \vspace{-1.5em} 
  \mycaption{The real (solid line) and imaginary (red dashed
  line) parts of the loop factor ratios $r_C={\cal C}^{(1)}/{\cal C}^{\rm LL}$ as a
  function of $q^2/m_e^2$ for vector (thick lines) and tensor (thin lines)
  lepton currents (these two lines overlap within the precision that can still be seen on the plot). The left plot is for $q^2$ negative (space-like momentum
  exchange), the right for $q^2$ positive (time-like momentum exchange). We
  take $m_\ell=m_e$ and $\mu=10$~GeV.} \label{fig:loopfactor}
\end{figure}

The $\chi N\to \chi N$ differential cross section $d\sigma/dE_d$ in the
leading log approximation is given in eq.~\ref{eq:vector:nucl}. Multiplying
it by $|{\cal C}_V^{(1)}/{\cal C}_V^{\rm LL}|^2$ one obtains the full 1-loop
prediction.  In fig.~\ref{fig:loopfactor} we show the value of ${\cal
C}_V^{(1)}/{\cal C}_V^{\rm LL}$  for $m_\ell=m_e$ and $\mu=10$~GeV. 
Note that above the pair production threshold, $q^2> 4 m_e^2$, it
develops an imaginary part, since electrons in the loop can go on-shell. The important thing for our purposes, though, is that ${\cal C}_V^{(1)}/{\cal
C}_V^{\rm LL}$ is a slowly varying function of $q^2$ and is of
$\mathcal{O}(1)$ in the range of space-like $q^2\sim -m_e^2$ of interest to
us, so that the LL approximation is quite precise. Even so, in the
numerical analysis in section~\ref{sec:results} we use the full 1-loop
results.

Let us next move to the tensor DM--lepton interaction, ${\cal L}_\ell=G (\bar
\chi \Gamma_\chi^{\mu\nu}\chi)(\bar \ell \sigma_{\mu\nu} \ell)$ with
$\Gamma_\chi^{\mu\nu}=(c_T+ic_{AT}\gamma_5)\sigma^{\mu\nu}$. The matrix
element for $\chi N\to \chi N$ scattering is
\be
\begin{split}
\mathcal{M}
 =& \,{\cal C}_T(\mu) \big(\bar u'_\chi \Gamma_\chi^{\mu\nu} u_\chi \big) \frac{q^\nu}{q^2}\langle N(p')| \sum_i Q_i \big(\bar q_i  \gamma_\mu q_i\big)|N(p)\rangle\\
 =& \,{\cal C}_T(\mu) \big(\bar u'_\chi \Gamma_\chi^{\mu\nu} u_\chi \big) \frac{q^\nu}{q^2} Z F(q) \big(\bar u'_N  \gamma_\mu u_N\big),
\end{split}
\ee
with the 1-loop factor in 
dimensional regularization (as before the pole is to be subtracted using $\overline{MS}$ scheme)
\be
{\cal C}_T^{(1)}(\mu)=-\frac{2\alpha_{\rm em}}{\pi} m_\ell GB_0(q^2,m^2,m^2)
\,.
\ee
Here $B_0(k^2, m^2,M^2)$ is a two-point scalar
Veltman-Passarino function. Explicit expressions for it can be found
e.g.\ in \cite{'tHooft:1978xw}.  In the leading log approximation the above expression becomes (in the $\overline{MS}$ scheme)
\be
{\cal C}_T^{\rm LL}(\mu)=\frac{2\alpha_{\rm em}}{\pi} m_\ell G\,
\log(m_\ell^2/\mu^2) \,.
\ee
In this limit the differential scattering cross section is given in
eq.~\ref{eq:tensor:nucl}, while the full 1-loop result is obtained by 
multiplying it with $|{\cal C}_T^{(1)}/{\cal C}_T^{\rm LL}|^2$. The
numerical value for the ratio ${\cal C}_T^{(1)}/{\cal C}_T^{\rm LL}$ is
shown in fig.~\ref{fig:loopfactor},  obtained using the {\tt LoopTools}
package~\cite{Hahn:1998yk}.

The scalar-type DM--lepton interaction ${\cal L}_{\rm eff}=G (\bar \chi
\Gamma_\chi\chi) (\bar \ell c_S^\ell \ell)$ with
$\Gamma_\chi=(c_S^\chi+ic_{P}^\chi\gamma_5)$, induces DM--quark interaction
through two-loop diagrams, see fig.~\ref{fig:loop-diagram}. This
contribution is relatively easy to compute in the limit of heavy leptons
using the operator product expansion. First one integrates out the heavy
leptons, thus matching onto the local dimension seven operator 
\be
{\cal L}_{\rm eff}=C_S(\mu)\frac{1}{m_\ell}(\bar \chi \Gamma_\chi \chi)
F_{\mu\nu}F^{\mu\nu}/e^2 \,,
\ee
where the Wilson coefficient is
\be
C_S=\frac{2}{3}\alpha_{\rm em}^2 G c_S^\ell \,.
\ee
This then enters a loop with two photons attached to the nucleon current. 
We evaluate this
1-loop diagram in the heavy nucleon limit, which gives for the
matrix element
\be
  \mathcal{M} =
  \frac{C_S}{16\pi^2} \, \frac{ 2\pi^2\kappa}{m_\ell} \,
  \big(\bar u_\chi'\Gamma_\chi u_\chi\big)
  Z^2 \tilde F(q)
  \big(\bar u_N'\tfrac{1}{2}(1+\gamma_0) u_N\big)  \,,
\ee
with $\kappa=\sqrt{2 m_N E_d}$ the recoil three-momentum of the nucleus, and
$\tilde F(q)$ the two-loop nuclear form factor. The resulting differential
cross section is given in eq.~\ref{scalar:two-loop:diff}. 

The derivation for scalar DM follows along the same lines. One matches onto
the dimension 6 operator
\be
{\cal L}_{\rm eff}=C_{S,5}(\mu)\frac{1}{m_\ell}(\chi^\dagger \chi) F_{\mu\nu}F^{\mu\nu}/e^2,
\ee
with the Wilson coefficient $C_{S,5}=\frac{2}{3}\alpha_{\rm em}^2 G_5
d_S^\ell$, which then gives a matrix element for $\chi N$ scattering
\be
  \mathcal{M} = \frac{C_{S,5}}{16 \pi^2} \, \frac{2\pi^2\kappa}{m_\ell}  Z^2 \tilde F(q) \, 
  \big(\bar u_N'\tfrac{1}{2}(1+\gamma_0) u_N\big)  \,.
\ee

\section{Derivation of the counting rate for WIMP--electron scattering}
\label{app:WE}

Using coordinate space Feynman rules we obtain for the matrix element for
WIMP scattering on an electron bound in the atomic shell with quantum
numbers $nlm$
\begin{align}
  \mathcal{M}^{rr'ss'}_{nlm} = G \int\!d^3x \, \psi_{nlm}(\vect{x}) \,
    e^{i (\vect{k} - \vect{k'} - \vect{p'}) \vect{x}} \,
    \bar{u}_\chi^{r'} \Gamma_\chi^\mu u_{\chi}^r \, \bar{u}_e^s \Gamma_{e\mu} u_e^s \,,
  \label{eq:ME-WEI}
\end{align}
Here, $\vect{k}$ and $\vect{k'}$ are the initial and final momenta of the WIMP,
and $\vect{p'}$ is the momentum of the electron in the final state. The
momentum of the initial state bound electron is determined by momentum
conservation: $\vect{p} \equiv \vect{k'} + \vect{p'} - \vect{k}$. Specializing
now to the axial vector case $\Gamma_\chi^\mu = \Gamma_e^\mu =
\gamma^\mu\gamma^5$ we obtain for non-relativistic $\chi$ and electrons%
\footnote{Using the hydrogen-like atom approximation we esimate the relativistic
corrections to be of order 20\% for electrons from the $1s$ shell of iodine and
smaller for the other shells.}
\be
\overline{|\mathcal{M}_{nlm}|^2} = \frac{1}{4}\sum_{rr'ss'}
|\mathcal{M}^{rr'ss'}_{nlm}|^2 = 48 m_e^2 m_\chi^2 G^2
|\psi_{nlm}(\vect{p})|^2 \,,
\ee
where the momentum space wave function $\psi_{nlm}(\vect{p})$ is defined by
\begin{align}
  \psi_{nlm}(\vect{p})
    =     \int\!d^3x \, \psi_{nlm}(\vect{x}) \, e^{-i \vect{p} \vect{x}}
  \equiv  \chi_{nl}(p) \, Y_{lm}(\theta,\phi) \,,
  \label{eq:psi-p}
\end{align}
with the normalization 
\be
\int \frac{d^3p}{(2\pi)^3} |\psi_{nlm}(\vect{p})|^2 = 1 \,.
\ee
For the differential cross section we have
\be
\frac{d\sigma_{nlm}^\wes}{dE_d} = \frac{\overline{|\mathcal{M}_{nlm}|^2}}
{32\pi E_\chi E_e v_{\chi e} |\vect{k} + \vect{p}|} \,.
\ee
Here, $E_\chi \approx m_\chi$, $E_e \approx m_e$ (using $E_B \ll m_e$), and
$v_{\chi e}$ is the relative velocity of the WIMP and the bound electron.
The event rate is obtained by summing over all shells and integrating over
the electron momenta in each shell:
\be\label{eq:WE-rate}
\frac{dR^\wes}{dE_d} = \frac{\rho_0 \eta_e}{m_\chi \rho_\mathrm{det}} 
\int \frac{d^3p}{(2\pi)^3} \int d^3v  \, v_{\chi e} f_\odot(\vect{v})
\sum_{nlm} \frac{d\sigma_{nlm}^\wns}{dE_d} \,.
\ee
The angular dependence of the wave function disappears 
due to the
orthogonality relation $\sum_m Y_{lm}(\theta,\phi) Y_{lm}^*(\theta,\phi) =
(2l+1)/4\pi$ for the spherical harmonics.
%
In the laboratory frame the electron and DM momentum form an angle $\theta$, so that 
$\cos\theta
= \vect{k} \vect{p} / k p$. For the integration over $\cos\theta$ from the
$d^3p$ integral we have to take into account eq.~\ref{eq:WE-Ed}, which holds
under the approximation $E_d \ll m_e \le E_e \ll m_\chi$ and implies
\begin{align}
  \cos\theta \approx 
    - \frac{1}{v} \left( \frac{E_d}{p} + \frac{p}{2 m_\chi} \right) \,.
  \label{eq:costh-avg}
\end{align}
The leading order corrections to this expression give the kinematically
available range for $\cos\theta$:
\begin{align}
  (\cos\theta)_{\rm max} - (\cos\theta)_{\rm min} \approx
    \frac{1}{m_\chi v p} \sqrt{2 m_e (E_d - E_B)(m_\chi^2 v^2 - 2 m_\chi E_d)} \,.
  \label{eq:costh-delta}
\end{align}
Then, $|\vect{k} + \vect{p}| \approx \sqrt{m_\chi^2 v^2 - 2 m_\chi E_d}$,
and eq.~\ref{eq:WE-rate} leads to the expression for the counting rate given
in eq.~\ref{eq:dRdE-WEI}. The approximations we have made in deriving the
formulas are accurate up to about 10\% for $m_\chi \sim 1$~GeV, but the
error decreases with increasing $m_\chi$, and we estimate an accuracy of
$\mathcal{O}(1\%)$ for $m_\chi \gtrsim 100$~GeV.

\section{Numerical evaluation of atomic matrix elements}
\label{app:wave-function}

In this appendix, we describe how we compute the radial momentum space
wave function $\chi(p)$ for WIMP--electron inelastic scattering and the
matrix elements $\bra{n'l'm'} e^{i (\vect{k} - \vect{k'}) \vect{x}}
\ket{nlm}$ defined in eq.~\ref{eq:ME-WAI-1} for WIMP--atom elastic and
inelastic scattering.

\bigskip

{\bf WIMP--electron scattering (\wes).} The momentum space radial wave
function $\chi_{nl}(p)$, see eq.~\ref{eq:psi-p}, required for the evaluation
of the event rate for \wes, is obtained by splitting the coordinate space
wave function $\psi_{nlm}(\vect{x})$ into its angular part
$Y_{lm}(\theta,\phi)$ and its radial part $R_{nl}(r)$, and computing
\begin{align}
  \chi_{nl}(p)\
    &= \frac{4\pi}{2l+1} \sum_m \psi_{nlm}(\vect{p}) \, Y_{lm}(\theta_p,\phi_p)
                                                                      \nonumber\\
    &= 2\pi \int\!dr \, r^2 R_{nl}(r) \int\!d(\cos\theta) \, P_l(\cos\theta) \,
       e^{i p r \cos\theta}
                                                                      \nonumber\\
    &= 4\pi i^l \int\!dr r^2 R_{nl}(r) j_l(pr) \,.
  \label{eq:chi-p}
\end{align}
Here, $\vect{p}$ is a momentum space vector with modulus $p$ and arbitrary
orientation $(\theta_p,\phi_p)$, and $P_l(\cos\theta)$ is a Legendre
polynomial.  In the second line, we have used the orthogonality of
the spherical harmonics, and in the third line, we have used Gegenbauer's
formula~\cite{Abramowitz:1964}, which relates the Fourier type integral over
a Legendre polynomial to the spherical Bessel function of the same degree.

The radial wave functions $R_{nl}(r)$ can be approximated by a linear
combination of so-called Slater type orbitals (STOs)~\cite{Bunge:1993a}:
\begin{align}
  R_{nl}(r) = \sum_k c_{nlk} \frac{(2 Z_{lk})^{n_{lk}+1/2}}{a_0^{3/2} \sqrt{(2n_{lk})!}}
              (r / a_0)^{n_{lk} - 1} \exp(-Z_{lk} r / a_0) \,.
  \label{eq:Slater}
\end{align}
Here, $a_0$ is the Bohr radius, and the parameters $c_{nlk}$, $n_{lk}$, and
$Z_{lk}$ are taken from~\cite{Bunge:1993a}. 

With $R_{nl}(r)$ given in the form of eq.~\ref{eq:Slater}, we can evaluate
\ref{eq:chi-p} analytically, which gives
\begin{align}
  \chi_{nl}(p) &= 
    \sum_k c_{nlk} \, 2^{-l + n_{lk}} \bigg( \frac{2\pi a_0}{Z_{lk}} \bigg)^{3/2}
    \bigg( \frac{i p a_0}{Z_{lk}} \bigg)^l
    \frac{(1 + n_{lk} + l)!}{\sqrt{(2n_{lk})!}} \,            \nonumber\\
  &\quad\times
    {}_2F_1 \bigg[ \tfrac{1}{2}(2 + l + n_{lk}), \tfrac{1}{2}(3 + l + n_{lk}),
                   \tfrac{3}{2} + l, -\bigg(\frac{p a_0}{Z_{lk}} \bigg)^2 \bigg] \,,
  \label{eq:chi-p-Slater}
\end{align}
with ${}_2F_1(a, b, c, x)$ being a hypergeometric function. 

\bigskip

{\bf WIMP--atom elastic scattering (\wae).} The matrix element for this case
can be written as
\begin{align}
  \sum_m \bra{nlm} e^{i (\vect{k} - \vect{k'}) \vect{x}} \ket{nlm}
  &= \sum_m \int \! dr \, d\Omega \, r^2 [R_{nl}(r)]^2 \, Y_{lm}^*(\theta,\phi) \,
    Y_{lm}(\theta,\phi) \, e^{i K r \cos\theta} \, \\
  &= (2l + 1) \int\!dr \, r^2 [R_{nl}(r)]^2 \, \frac{\sin Kr}{Kr}
  \label{eq:ME-numerics-1}
\end{align}
with the abbreviation $K \equiv |\vect{K}| \equiv |\vect{k} - \vect{k'}|$. This
integral has the form of a Fourier Sine Transform, and can be evaluated
efficiently using the Fast Fourier Transform (FFT) algorithm. For $R_{nl}(r)$,
we use again the expansion eq.~\ref{eq:Slater}, with the coefficients taken
from~\cite{Bunge:1993a}.

\bigskip

\begin{figure}
  \begin{center}
    \includegraphics[width=\textwidth]{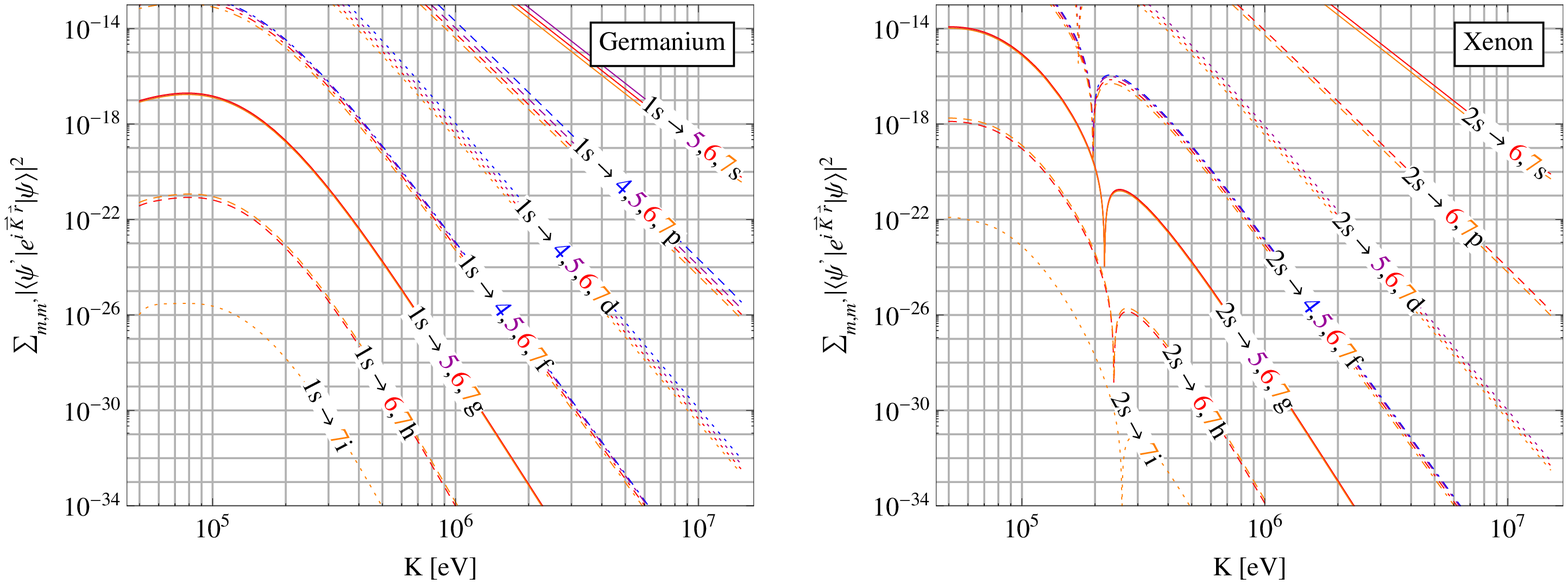}
  \end{center}
  \vspace{-1.5em}
  \caption{A few examples for \wai\ matrix elements of germanium and xenon. The
    plot shows transitions from the $1s$ resp.\ $2s$ orbitals to some of the
    lowest unoccupied states.}
  \label{fig:me-wai}
\end{figure}

{\bf WIMP--atom inelastic scattering (\wai).} Here the numerical evaluation of
the atomic matrix elements is slightly more involved than for \wae\ because now
$\psi_{n'l'm'}(\vect{x}) \neq \psi_{nlm}(\vect{x})$.  We expand the factor
$e^{i \vect{K} \vect{x}}$ in eq.~\ref{eq:ME-WAI-1} in spherical
harmonics~\cite{Jackson:EDyn} and rewrite the angular integral over a product
of three spherical harmonics in terms of the Wigner-3j
symbols~\cite{Sakurai:QM}. This gives
\begin{align}
  \bra{n'l'm'} e^{i \vect{K} \vect{x}} \ket{nlm} &=
    4\pi \int\!dr \, r^2 R_{nl}(r) \, R_{n'l'}(r)
    \sum_{L,M} j_L(Kr) \, Y_{LM}(\theta_K, \phi_K) \nonumber\\
  &\quad\times
    \frac{(-1)^m}{\sqrt{4\pi}} \sqrt{(2l+1)(2l'+1)(2L+1)}
    \begin{pmatrix}
      l  & l' & L \\
      0  & 0  & 0
    \end{pmatrix}
    \begin{pmatrix}
      l & l' & L \\
      m & m' & M
    \end{pmatrix} \,,
  \label{eq:ME-WAI-2}
\end{align}
where $j_L$ denotes a spherical Bessel function of the first kind, and
$\theta_K$, $\phi_K$ are the angular components of $\vect{K}$. To compute the
cross section, we need the expression
\begin{align}
  \sum_{mm'} \Big| \bra{n'l'm'} e^{i \vect{K} \vect{x}} \ket{nlm} \Big|^2
  &= (2l+1)(2l'+1) \sum_L (2L+1) \bigg[ \begin{pmatrix}
                                          l & l' & L \\
                                          0 & 0  & 0
                                        \end{pmatrix} \bigg]^2 \nonumber\\
  &\quad\times
     \bigg[ \int\!dr \, r^2 R_{nl}(r) \, R_{n'l'}(r) j_L(Kr) \bigg]^2 \,.
  \label{eq:ME-WAI-3}
\end{align}
Here, we have used the symmetry and orthogonality relations of the Wigner-3j
symbols and of the spherical harmonics.  The expression in the second set of
square brackets has the form of a spherical Bessel transform, which we
evaluate by using an algorithm due to Sharafeddin et
al.~\cite{Sharafeddin:1992a}, based on rewriting the spherical Bessel
function as a finite Fourier series, thus converting the integral to a sum
of Fourier sine and Fourier cosine transforms. The initial state wave
functions $R_{nl}$ are again given by eq.~\ref{eq:Slater} and
ref.~\cite{Bunge:1993a}, while for the final state wave functions
$R_{n'l'}$, we use the hydrogen-like approximation (with an effective charge
$Z=3$ due to screening of the nuclear charge by inner electrons) since
accurate tabulated wave functions for excited atoms were not available.
Also, we consider only transitions from the $1s$, $2s$, $3s$ levels to the
first two unoccupied $s$ shells; we have checked numerically that these
transitions are the most important ones. Fig.~\ref{fig:me-wai} shows some of
the \wai\ matrix elements for germanium and xenon. We find that for
germanium (xenon) transitions from the $1s$ ($2s$) orbital dominate. Our
approximations should correctly reproduce the qualitative behavior of the
matrix elements at large momentum transfer, and should lead to a good order
of magnitude estimate of the \wai\ event rate.

\section{On the equilibrium of DM capture and annihilations in the Sun}
\label{app:equil}

In the calculation of the neutrino flux from DM annihilations in the Sun we
have assumed that WIMP captures and annihilations are in equilibrium, which
makes the result independent of the DM annihilation cross section $\langle
\sigma_\mathrm{ann} v\rangle$.  Here we comment on the validity of this
assumption. 
Let us first estimate the cut-off scale $\Lambda$ for the effective theory
description of the DM--lepton coupling. For the two examples of $V\otimes V$ and
$A\otimes A$ couplings, the neutrino bounds are of order $\sigma_{\chi e}^0 \sim
10^{-43}\, \rm cm^2$ and $10^{-38}\, \rm cm^2$, respectively, see
figs.~\ref{fig:msigma-V} and \ref{fig:msigma-A}. From eq.~\ref{eq:sig0ferm}
we can estimate the corresponding cut-off scales as $\Lambda_V \sim 100$~GeV
and $\Lambda_A \sim 10$~GeV, 
where we took coupling constants $c_i^\chi$ to be of order
$\mathcal{O}(1)$. In DM annihilations the four-momentum transfer
squared is of order $m_\chi^2$.
For $m_\chi \sim 10$~GeV,
relevant for \wns, the WIMP annihilations may then also be 
described by effective field theory. 
Using effective interactions in eq. \ref{eq:4fermi} (extending them to neutrinos),
we find
\be\label{eq:sigannV}
\text{Vector:}\quad
\langle \sigma_\mathrm{ann} v\rangle \sim \frac{G^2 m_\chi^2}{\pi} =
\sigma_{\chi e}^0 \frac{m_\chi^2}{m_e^2} \sim 
10^{-24} \, {\rm cm^3\,s^{-1}} 
\bfrac{\sigma_{\chi e}^0}{10^{-43}\,{\rm cm^2}}
\bfrac{m_\chi}{10\,\rm GeV}^2 \,.
\ee
In the \wes\ case, however, the effective theory typically cannot be applied
since the momentum transfer for annihilations is above the cut-off scale.
Therefore, in general we cannot make model independent statements about
$\langle \sigma_\mathrm{ann} v\rangle$ without specifying the UV completion
of the effective $\chi\ell$ vertex. 
An order of magnitude estimate
can still be obtained from dimensional analysis as 
\be\label{eq:sigannA}
\text{Axial:}\qquad
\langle \sigma_\mathrm{ann} v\rangle \sim \frac{g^4}{m_\chi^2} \sim 
10^{-21} \, {\rm cm^3\,s^{-1}} \times g^4
\bfrac{m_\chi}{100\,\rm GeV}^{-2} \,,
\ee
with $g$ a typical coupling constant between leptons and the dark sector.

\begin{figure}[t]
\centering 
\includegraphics[width=0.6\textwidth]{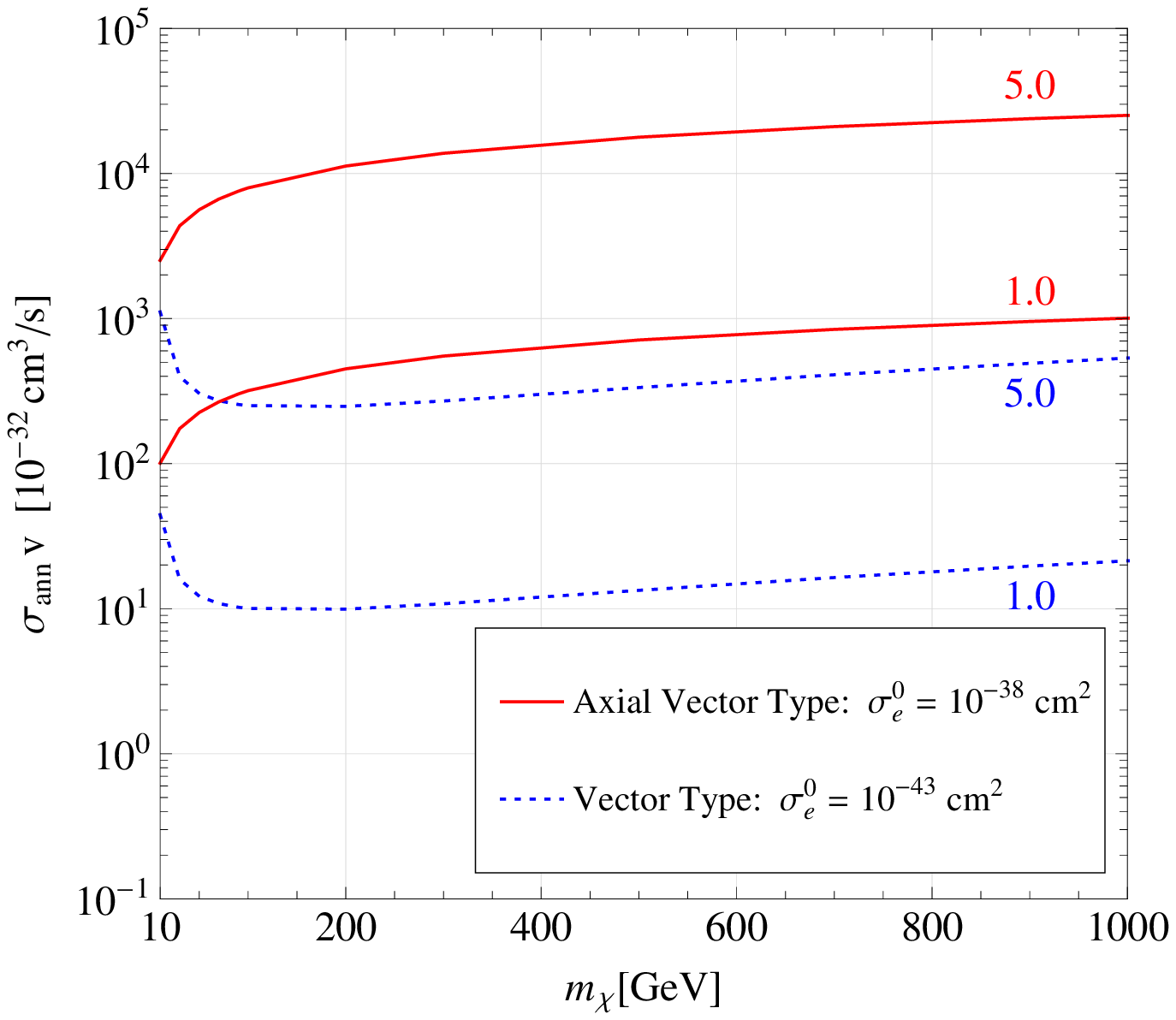}
  \mycaption{\label{fig:equil} Contours of $t_\odot/\tau = 5$ and
  $t_\odot/\tau = 1$, see eq.~\ref{eq:equil}. For the case of vector (axial
  vector) coupling we have used a scattering cross section of $\sigma_{\chi
  e}^0 = G^2 m_e^2/\pi = 10^{-43} (10^{-38})\,\rm cm^2$, motivated by the
  results of the Super-Kamiokande bound. For values of $\langle
  \sigma_\mathrm{ann} v\rangle$ above the curve for $t_\odot/\tau = 5$, WIMP
  capture and annihilations are in equilibrium in the Sun.}
\end{figure}

Equilibrium of WIMP capture and annihilations is obtained if 
$\tanh^2(t_\odot/\tau)$ is close to 1, see eq.~\ref{eq:equil}.
Fig.~\ref{fig:equil} shows the values of $\langle \sigma_\mathrm{ann}
v\rangle$ for which $t_\odot/\tau = 1$ and $5$ as a function of $m_\chi$.
The values of scattering cross sections
$\sigma_{\chi e}^0$ for $V\otimes V$ and $A\otimes A$ Lorentz structures were chosen to be 
above (but close to) the Super-Kamiokande bounds shown in
figs.~\ref{fig:msigma-V} and \ref{fig:msigma-A}. 
Since $\tanh^2 x \approx 1$
for $x\gtrsim 5$, WIMP capture and annihilations are in equilibrium in the
Sun for values of $\langle \sigma_\mathrm{ann} v\rangle$ above the curve for
$t_\odot/\tau = 5$. Comparing eqs.~\ref{eq:sigannV} and \ref{eq:sigannA}
with the ranges shown in the figure we conclude that the assumption of
equilibrium is very well justified in the cases of our interest.

\bibliographystyle{apsrev}
\bibliography{./dama-el}

\end{document}